\documentclass{aa}

\usepackage{graphicx}
\usepackage{txfonts}
\usepackage{amsmath}
\usepackage[]{hyperref}

\hypersetup{
    colorlinks = true,
    linkcolor = {blue}, 
    citecolor = {blue},
    urlcolor = {blue}
}

\usepackage{natbib}
\usepackage{color}
\usepackage{float}
\usepackage{xspace} 
\usepackage{multirow}

\definecolor{lightblue}{rgb}{0.2, 0.6, 1}
\definecolor{darkred}{rgb}{0.81, 0.09, 0.13}

\newcommand{\citesoftware}[2]{{\href{#2}{\textsf{\color{lightblue} #1}}}\xspace}

\newcommand{\spitzer}{\textsl{Spitzer}\xspace}
\newcommand{\kepler}{\textsl{Kepler}\xspace}

\newcommand{\kelp}{\citesoftware{kelp}{https://github.com/bmorris3/kelp}}
\newcommand{\hml}{\ensuremath{h_{m\ell}}\xspace}

\begin{document}

\title{Observations of scattered light from exoplanet atmospheres}
\author{Brett M.~Morris\inst{1}
\and Kevin Heng\inst{2, 3, 4}
\and Daniel Kitzmann\inst{5}
}

\institute{Space Telescope Science Institute, 3700 San Martin Dr, Baltimore, MD 21218, USA \and 
Ludwig Maximilian University, University Observatory Munich, Scheinerstrasse 1, Munich D-81679, Germany \and
University of Warwick, Department of Physics, Astronomy \& Astrophysics Group, Coventry CV4 7AL, United Kingdom \and
University of Bern, ARTORG Center for Biomedical Engineering Research, Murtenstrasse 50, CH-3008, Bern, Switzerland \and
Center for Space and Habitability, University of Bern, Gesellschaftsstrasse 6, CH-3012 Bern, Switzerland}

   \date{Submitted April 21, 2022; accepted January 24, 2024}
 
  \abstract
   {Optical phase curves of hot Jupiters can reveal global scattering properties. We implement a Bayesian inference framework for optical phase curves with flux contributions from: reflected light from a potentially inhomogeneous atmosphere, thermal emission, ellipsoidal variations, Doppler beaming, and stellar rotation via a Gaussian process in the time domain. We probe for atmospheric homogeneity and time-variability using the reflected light inferences for highly precise \kepler light curves of five hot Jupiters. We also investigate the scattering properties which constrain the most likely condensates in the inhomogeneous atmospheres. Cross validation prefers inhomogeneous albedo distributions for Kepler-7 b and Kepler-41 b, and a weak preference for inhomogeneity for KOI-13 b. None of the five planets exhibit significant variations in geometric albedo on one-year timescales, in agreement with theoretical expectations. We show that analytic reflected light phase curves with isotropic multiple scattering are in excellent agreement with full Rayleigh multiple scattering calculations, allowing for accelerated and analytic inference. In a case study of Kepler-41 b, we identify perovskite, forsterite, and enstatite as possible scattering species consistent with the reflected light phase curves, with condensate particle radii in the range $0.01-0.1$ µm. }


   \maketitle

\section{Introduction}

Optical phase curves of exoplanets can constrain properties of planetary atmospheres \citep[see reviews by][]{Deming2017,Parmentier2018}. The phase curve encodes degenerate information about the temperature of the photosphere or surface, the reflectance or scattering properties of the photosphere or surface, and the effects of the planet on the host star such as ellipsoidal variations and Doppler beaming.

The amplitudes and shapes of reflected light phase curves are functions of fundamental properties of the photosphere or scattering surface \citep{Cowan2011,Demory2013,Hu2015,Shporer2015,Munoz2015,Jansen2018,Farr2018,Mayorga2020,Luger2021,Fraine2021}. These properties include the single-scattering albedo as a function of longitude, and the asymmetry of scattering events (forward or backward scattering). In general these quantities are also functions of wavelength, and broad-optical phase curves constrain the bandpass-integrated quantities. From these scattering parameters, the geometric albedo and integral phase function can be computed.

Reflected light phase curves of exoplanet atmospheres are contaminated by thermal emission, ellipsoidal variations, and Doppler beaming \citep[e.g.:][]{Demory2011b,Demory2013,Shporer2014,Esteves2015}. The precise ratio of reflected light to thermal emission is often poorly constrained by optical observations alone, since the contribution from thermal emission often peaks at a similar orbital phase to the orbital phase of maximal reflected light for a homogeneous reflector \citep{Demory2013}. For short-period planets, the hottest point on the dayside of a planet with strong atmospheric circulation may not be at the substellar point, which can sometimes be equally described without circulation but with inhomogeneous cloud coverage, when measured with photometry in the optical and near-infrared. The ellipsoidal and Doppler contributions to the phase curve are constrained by the orbital period of the planet with uncertain amplitudes. Due to the uncertain mixture of these contributions in a given phase curve, all these effects must be considered simultaneously.

The joint inference required for studying planets in reflected light is strongly motivated by the planetary constraints revealed by reflected light. Phase curves can directly constrain two optical properties of condensates in the atmosphere, namely the single scattering albedo and scattering asymmetry parameter, which are linked to the species of condensates responsible for the scattering \citep{Kitzmann2018}. Furthermore, with the analytic reflection model in \citet{Heng2021}, the shape of the reflected light phase curve can be linked to a scattering phase function. Common phase functions used in the exoplanet literature include Rayleigh \citep{Hubbard2001}, isotropic \citep{Demory2013}, Henyey-Greenstein \citep{Sudarsky2000,deKok2012,Robinson2017} and double Henyey-Greenstein \citep{Heng2021b,Adams2022} for example. The choice of a particular planetary scattering phase function can be directly linked to the scattering phase function of condensates, given a particle composition and size as well as cloud coverage. The Solar System planets offer guideposts with more complex scattering phase functions. \citet[][see Figure~5]{Mayorga2016} and \citet[][see Figures~4, 5]{Dyudina2016} found that flyby observations of Jupiter are inconsistent with pure isotropic or Rayleigh scattering phase functions\footnote{Much older observations like those, for example, of Venus in \citet[][Figure~3]{Horak1950} also show the non-ideal scattering phase function behavior in the Solar System.}. These precision Solar System observations and theoretical advances indicate that the scattering phase function can and should be inferred from the phase curves themselves.

Stellar rotation and instrumental systematics generate correlated signals in the aperture photometry of exoplanet host stars. A Bayesian retrieval framework for exoplanet phase curve parameters must also account for the uncertainty in the detrending technique used to account for stellar rotation and instrumental systematics. Gaussian process (GP) regression is an effective and efficient technique for removing these time-correlated signals, and propagating the uncertainties in the kernel hyperparameters that describe the correlation in time into the atmospheric parameters retrieved for a system \citep[see e.g.:][]{Angus2018}. 

Markov Chain Monte Carlo (MCMC) methods are commonly used in astrophysics to estimate posterior distributions for uncertain parameters from measurements \cite[see review by][]{Sharma2017}. Recently, gradient-based inference techniques are becoming more widely used in photometric analyses of exoplanets, such as Hamiltonian Monte Carlo (HMC) and its popular variant, the No U-Turn Sampler \citep[NUTS; see for example:][]{exoplanet:luger18,exoplanet:agol20,Colon2020,Price-Whelan2020,Stefansson2020,Dalba2021,Daylan2021,Agol2021,VanEylen2021,Foreman-Mackey2021}. NUTS is especially useful for efficiently sampling high dimensional and degenerate posterior distributions \citep{Betancourt2017}. Many of the probabilistic programming frameworks which implement HMC methods also provide methods for Leave-One-Out cross-validation, which can be used to compare and select among a suite of models \citep{}. This approach has been applied more recently in astronomical contexts \citep{Welbanks2023,Gandhi2023,Challener2023}.

In this work, we develop a Bayesian inference framework for determining reflected light properties of hot Jupiters observed with \kepler. We outline the photometric model and its sampler in Section~\ref{sec:methods}. We present the results of the photometric analysis and a predictive accuracy/model comparison technique in Section~\ref{sec:results}, including assessments of inhomogeneity in exoplanet albedos, time variability, and inference of the single scattering phase function. We interpret these results in terms of potential condensates and observational biases in Section~\ref{sec:interpretation}. We discuss the implications in Section~\ref{sec:discussion}, and conclude in Section~\ref{sec:conclusion}.

\section{Methods} \label{sec:methods}

In this section we outline the numerical methods for sampling from the posterior distributions for the single-scattering albedos and scattering asymmetry parameters of exoplanet atmospheres with optical phase curve photometry. The source code is freely available in a package called \kelp\footnote{\url{https://github.com/bmorris3/kelp}}. For computational efficiency, the phase curve models have been implemented in \textsf{JAX} \citep{jax2018github}, a transpiled framework in Python. All phase curve parameter priors are listed in Table~\ref{tab:priors}.

\begin{table}
    \centering
    \setstretch{1.2}
    \begin{tabular}{lcr}
        Parameter & Prior dist./value & Interval  \\\hline
        \multicolumn{3}{l}{\it Reflected light}\\
        $\omega_0$ & 0 & -- \\
        $\omega^\prime$ & $\mathcal{U}$ & [0, 1] \\
        $g$ & $\mathcal{N}(0, 0.01)$ & $[-1, 1]$ \\ 
        $A_g$ & $\mathcal{U}$ (for inh. models$^\star)$ & $[0, 0.6]$ \\
        $x_1^\prime = \sin(x_1)$ & $\mathcal{U}$ & $[-1, 0]$\\
        $x_2^\prime = \sin(x_2)$ & $\mathcal{U}$ & $[0, 1]$\\\\
        \multicolumn{3}{l}{\it Thermal emission}\\
        $C_{11}$ & $\mathcal{N}(0, 0.1)$ & $[0, 0.2]$ \\
        $\alpha$ & 0.6 & -- \\ 
        $\omega_\mathrm{drag}$ & 4.5 & --  \\ 
        $f$ & $\mathcal{N}(2^{-0.5}, 0.01)$ & [0.67, 0.74]  \\
        $\Delta\phi$ & $0^\circ$ & -- \\\\
        \multicolumn{3}{l}{\it Stellar artifacts}\\
        Ellipsoidal amp. & $\mathcal{N}(A_\mathrm{ellip}, 0.25A_\mathrm{ellip})$ & $[0, 100]$ ppm \\ 
        Doppler amp. & $\mathcal{N}(A_\mathrm{Doppler}, 0.25A_\mathrm{Doppler})$ & $[0, 50]$ ppm \\\\
        \multicolumn{3}{l}{\it GP \& uncertainty}\\
        $\sigma_\mathrm{GP}$ & $\mathcal{N}(\sigma_f, 0.1\sigma_f)$ & $[0, 10\sigma_f]$ ppm \\
        $\rho_\mathrm{GP}$ & 30 d & -- \\
        Jitter & $\mathcal{U}$ & $[0, 10^3]$ ppm\\
    \end{tabular}
    \caption{Prior distributions or values adopted for the reflected light from \citet{Heng2021} and thermal emission phase curves from \citet{Morris2021_hml}, as well as parameters for stellar and instrumental artifacts. $^\star$The inhomogeneous reflector model defines $A_g$ as a free parameter, whereas the homogeneous reflector model derives a deterministic $A_g$ from $\omega,\; g,$ and the scattering phase function. }
    \label{tab:priors}
\end{table}

\subsection{Photometry and system parameters}

\begin{figure}
    \centering
    \includegraphics[width=\columnwidth]{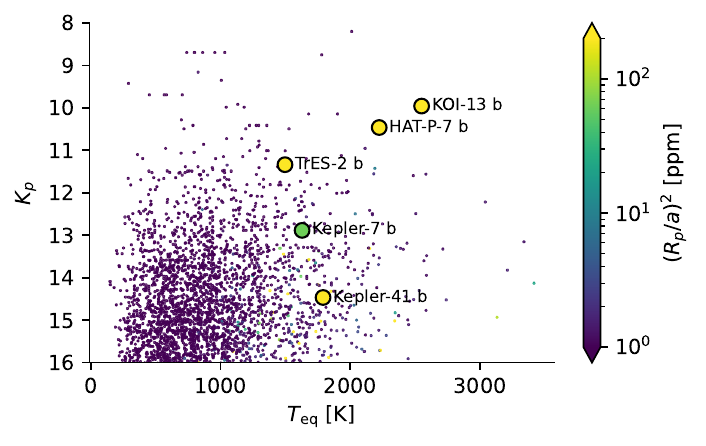}
    \caption{Host star \kepler magnitudes and their planetary equilibrium temperatures for the five planets we consider (large points). The color of each point indicates the amplitude of the reflected light phase curve assuming $A_g = 1$ (upper limit). }
    \label{fig:targets}
\end{figure}

In this work, we focus our attention on planets with equilibrium temperatures $T_{\rm eq} \gtrsim 1500$ K, as the daysides of these planets are likely to transition from cloudy to cloud-free as temperature increases above the stability limits of condensates near the substellar longitude. We selected five planets spanning a range of equilibrium temperatures above 1500 K with geometries favorable for measuring reflected light. The sample of selected planets is shown in Figure~\ref{fig:targets}.

We fix the orbital period, mid-transit time, the stellar mass, planetary mass, planetary radius, semimajor axis, stellar density, impact parameter, and stellar temperature to literature values curated by the NASA Exoplanet Archive in the \textsf{PSCompPars} table.

We retrieve each long-cadence light curve from MAST as Simple Aperture Photomery (SAP) fluxes. We detrend the unmasked light curves with cotrending basis vectors (CBVs) also downloaded from MAST\footnote{\url{https://mast.stsci.edu/}} via \textsf{lightkurve} \citep{LightkurveCollaboration2018}, extracting the first eight basis vectors and applying single-scale CBV correction with the L2-norm regularization penalty $\alpha = 10^{-4}$.

\subsection{Reflected light}

The reflected light phase curve is computed with the {\it ab initio} solutions for any reflection law by \citet{Heng2021}. We test models with both homogeneous and inhomogeneous planetary atmospheres, where the latter is a generalization of the piecewise-Lambertian model of \citet{Hu2015}. The inhomogeneous sphere has a dark region with single scattering albedo $0 \leq \omega_0 \leq 1$ which is surrounded by a brighter region $0 < \omega \equiv \omega_0 + \omega^\prime < 1$. The less reflective region is bounded by longitudes $-\pi/2 < x_1 < x_2 < \pi/2$. The geometric albedo is the last free parameter, which completes the set of parameters necessary to derive the scattering asymmetry parameter $-1 < g < 1$.

We apply a prior to the scattering asymmetry parameter $g \sim \mathcal{N}(0, 0.01)$ since the approximations in the reflected light model are most accurate for strongly asymmetric phase functions when $g$ is closer to zero.
The reflected light model for an inhomogeneous atmosphere is given by
\begin{equation}
    F_\mathrm{reflect} = \left(\frac{R_p}{a}\right)^2 A_g \Psi(\omega_0, \omega, g, x_1, x_2),
\end{equation}
where $R_p$ is the planet radius, $a$ is the semimajor axis, $A_g$ is the geometric albedo, and the functional form of the integral phase function $\Psi$ is stated in equations (40) and (42) of \citet{Heng2021}. When $\omega_0 = \omega$, the inhomogeneous sphere model reduces to a homogeneous sphere. 

The less-reflective substellar region is bounded by longitudes $x_1$ and $x_2$, defined relative to the substellar longitude. To sample longitudes uniformly in the reference frame of a distant observer, we sample $x_1 = \sin^{-1}(x_1^\prime)$ and $x_2 = \sin^{-1}(x_2^\prime)$, with $x_1^\prime \sim \mathcal{U}(-1, 0)$ and $x_2^\prime \sim \mathcal{U}(0, 1)$.

\subsubsection{Single scattering albedo}

The two-albedo inhomogeneous model was inspired by the phase curve of Kepler-7 b \citep{Hu2015}. The physical motivation for the two albedos used for Kepler-7 b comes from theoretical expectations that cloudy regions have high albedos in comparison with low-albedo cloud-free regions. The temperature near the substellar longitude in these atmospheres is likely too hot for condensates to persist, though at some longitude closer to the limb, the temperatures may dip below the stability curves of highly refractory species. The inhomogeneous model captures this behavior with a dark substellar region with albedo $\omega_0$, bounded by freely-varying longitudes to the west and east where the albedo increases to $\omega$.

Phase curve analyses with two-albedo models usually infer that the substellar regions have near-zero albedo \citep{Hu2015, Adams2022}. We confirm that fits to the phase curves with a uniform, uninformative prior on $\omega_0$, and we find that all phase curves in this work are consistent with $\omega_0 \approx 0$. In all subsequent fits, including all results presented in this work, we simplify the inhomogeneous model parameterization by fixing $\omega_0=0$ in the substellar region. In addition to improving fit convergence, fixing $\omega_0$ simplifies the interpretation of the results in Section~\ref{sec:interpretation}, since we can cleanly assume the region between $x_1$ and $x_2$ is cloud-free. 

We note that the single-scattering albedo constrained in this work is defined mathematically by \citet{Heng2021}, and its application to \kepler phase curves has some subtleties. The single-scattering albedo is defined for a particular wavelength. Since \kepler observations span a red-optical bandpass, the single-scattering albedos reported here are flux-weighted in the \kepler bandpass. The albedo measured here is also a weighted mean over the single-scattering albedo of the gas as well as the single-scattering albedo of the clouds.

\subsubsection{Scattering phase function}

\begin{figure}
    \centering
    \includegraphics[width=\columnwidth]{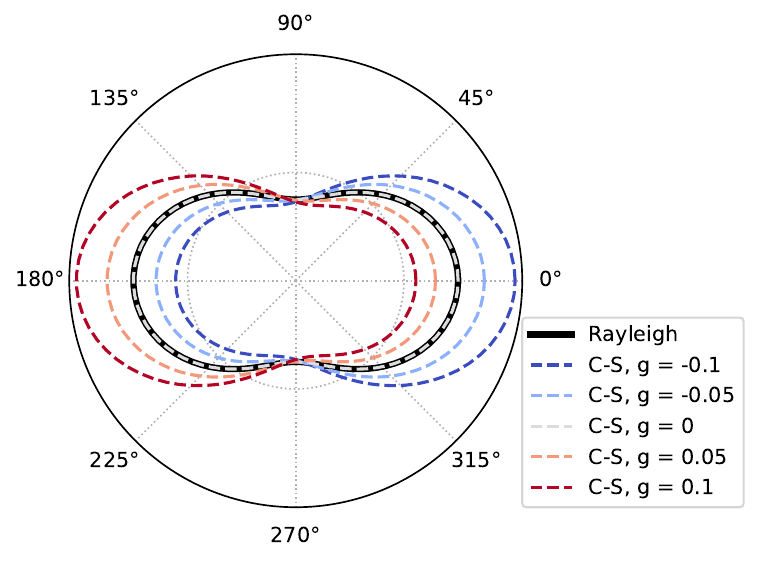}
    \caption{The scattering phase function proposed by \citet[CS,][]{Cornette1992}, which reduces to Rayleigh when $g = 0$. }
    \label{fig:cornette}
\end{figure}

The phase curve formulae from \citet{Heng2021} allow users to trivially substitute different scattering phase functions for the exoplanet atmosphere. In principle, the choice of scattering phase function maps onto assumptions about the particle size in the planet's atmosphere, the type of condensate (refractive indices), cloudiness, and the proportion of single-to-multiple scattering. Three simple scattering phase functions are considered in this work: (1) isotropic, which is convenient to implement, though practically one should not expect phase functions from single scattering events to be isotropic; (2) Rayleigh, which is expected for clear or cloudy atmospheres with particles that are much smaller than the wavelength of light considered \citep{Rayleigh1871, Rayleigh1899}; and (3) the \citet{Cornette1992} scattering phase function,
\begin{equation}
    P = \frac{3}{2} \frac{1 - g^2}{2 + g^2} \frac{1 + \mu^2}{(1 + g^2 - 2g\mu)^{3/2}},
\end{equation}
shown in Figure~\ref{fig:cornette}. The latter is a generalization of the Rayleigh scattering phase function with additional freedom for stronger forward or backward scattering, parameterized by the scattering asymmetry parameter $g$. In the limit of $g = 0$, the Cornette-Shanks scattering phase function correctly reduces to Rayleigh instead of isotropic scattering.

\subsection{Thermal emission}

We model thermal emission of each exoplanet in the optical with the thermal emission model defined in \citet{Morris2021_hml}. The so-called \hml basis describes the temperature map of the exoplanet using generalized spherical harmonics \citep[parabolic cylinder functions,][]{Heng2014}. There are two parameters which define the strength of the chevron shape of the hotspot feature on the dayside, which we fix to the values compatible with both GCM temperature maps and \spitzer phase curves, $\alpha=0.6$ and $\omega_\mathrm{drag}=4.5$. Details and validation for this approach are provided in \citet{Morris2021_hml}. 

We sample for two free parameters that define the atmosphere's temperature map, which produces thermal emission. The first is the power in the $m = \ell = 1$ spherical harmonic term, $C_{11}$, which determines the contrast between the dayside and nightside temperatures. Expected values for $C_{11}$ are of order $0.1$, so we draw from a normal distribution $C_{11} \sim \mathcal{N}(0, 0.1)$, with lower and upper limits at zero and 0.2. The lower limit corresponds to the limiting case of perfect heat redistribution from the dayside to the nightside, and the upper limit prevents extreme day/night contrasts. 

The second free parameter is $f$, which is sometimes called the greenhouse factor. If we write the temperature map of the planet as a perturbation around a mean temperature $\bar{T}$, which is similar to the equilibrium temperature,
\begin{equation}
\bar{T} = f \; T_\star \sqrt{R_\star/a},
\end{equation}
$f$ is related to the atmosphere's efficiency at converting incoming stellar radiation into thermal emission (see Sections 2 and 3 of \citealt{Morris2021_hml} for more discussion). We sample within a small range about the expected value $f \sim \mathcal{N}(2^{-1/2}, 0.01)$.

We find that the prior's upper limit $C_{11} < 0.2$ is necessary for fits to optical phase curves in this parameterization. Primarily, $f$ sets the global mean temperature and $C_{11}$ sets the day/night contrast. Since the great majority of the flux from thermal emission  is detected at full phase ($\xi = 0$), the hemisphere-averaged dayside temperature would be the best-constrained quantity to measure from an thermal phase curve in the optical. The dayside temperature in this parameterization results from a combination of $f$ and $C_{11}$, where these terms are somewhat degenerate -- a higher global mean temperature and smaller contrast can produce the same dayside temperature as a lower global mean temperature and a higher contrast. In infrared phase curves, as in \citet{Morris2021_hml}, this degeneracy can be avoided because the nightside thermal emission can be clearly detected. The nightside emission is negligible in the optical \kepler bandpass, so an upper limit on $C_{11}$ prevents sampling extreme solutions with hot daysides, cool nightsides, and low global mean temperatures.

We assume zero hotspot offset for each planet, $\Delta\phi=0$. \spitzer phase curve analysis in \citep{Bell2021} and \citet[][see their Figure 8]{Morris2021_hml} showed that the hotspot offset is often small for planets with $1000 \mathrm{~K} < T_\mathrm{eq} < 2500$ K, which encompasses all planets in this sample. Practically, assuming zero hotspot offset also provides the maximum contamination of the reflected light signal from thermal emission, and allowing $C_{11}$ to vary allows us to explore the degeneracy between thermal emission and reflected light at secondary eclipse. Fixing $\Delta\phi=0$ therefore implies that the geometric albedos that we report may be lower limits, since a non-zero hotspot offset will increase the contribution of reflected light to the phase curve amplitude.

\subsection{Ellipsoidal variations and Doppler beaming}

We use the following relations to estimate the amplitudes of the ellipsoidal variations and Doppler beaming respectively \citep[see e.g.:][and references therein]{Shporer2014}:
\begin{eqnarray}
    A_\mathrm{ellip} \mathrm{~[ppm]} &\approx& \frac{\alpha_\mathrm{ellip}}{0.077} \frac{M_p}{M_J} \left(\frac{R_\star}{R_\odot}\right)^{3} \left(\frac{M_\star}{M_\odot}\right)^{-2} \left(\frac{P}{\mathrm{day}}\right)^{-2}\\
    A_\mathrm{Doppler} \mathrm{~[ppm]} &\approx& \frac{\alpha_\mathrm{Doppler}}{0.37} \frac{M_p}{M_J} \left(\frac{M_\star}{M_\odot}\right)^{-2/3} \left(\frac{P}{\mathrm{day}}\right)^{-1/3}
\end{eqnarray}
where each $\alpha$ coefficient is a factor of order unity.

The flux from ellipsoidal variations has the form
\begin{equation}
    F_\mathrm{ellip} = A_\mathrm{ellip} \left(1 - \cos{4\pi(\phi_\mathrm{orb}} - 0.5)\right),
\end{equation}
where the orbital phase $\phi_\mathrm{orb}$ is normalized to vary from zero at mid-transit to $0.5$ at secondary eclipse. 

The Doppler beaming model has the form
\begin{equation}
    F_\mathrm{Doppler} = A_\mathrm{Doppler} \sin{2\pi\phi_\mathrm{orb}},
\end{equation}
and generally has a smaller amplitude than the ellipsoidal variations.

\subsection{Eclipse model}

The secondary eclipse model for each planet $\lambda^e$ is the occultation defined by \citet{exoplanet:agol20} as implemented in \textsf{exoplanet} \citep{Foreman-Mackey2021}. We normalize $\lambda^e$ such that it is unity out-of-eclipse and zero between second and third contacts, so that we can multiply the planetary contributions to the phase curve by this function. 

\subsection{Flux contamination from nearby stars} \label{sec:contamination}

In \kepler observations, the flux from the target star may be diluted by flux from neighboring stars which falls within the photometric aperture of the target star. For each target except KOI-13, we adopt the Simple Aperture Photometry (SAP) crowding factor \textsf{CROWDSAP} in the \kepler and TESS input catalogs as the fraction of flux in the aperture from the target star.
For KOI-13, we adopt the crowding factor due to flux from the other components in the hierarchical triple measured by \citet{Shporer2014}.

\subsection{Composite mean model}

The complete phase curve model is given by
\begin{equation}
    \frac{F_p}{F_\star} = \left( \lambda^e (F_\mathrm{reflect} + F_\mathrm{thermal}) + F_\mathrm{ellip} + F_\mathrm{Doppler} \right) / \delta_F,
\end{equation}
where $\delta_F$ is the crowding factor as described in Section~\ref{sec:contamination}.

\subsection{Gaussian process regression} \label{sec:gp}

We incorporate a Gaussian process regression step into the phase curve analysis using the Mat{\'e}rn 3/2 kernel implemented in \textsf{celerite2} \citep{celerite2}. We use a fixed timescale $\rho_\mathrm{GP} = 30$ days and unknown standard deviation of the process $\sigma_\mathrm{GP}$. The fixed $\rho_\mathrm{GP}$ timescale is about ten-times longer than the phase curve variations ($P<3$ d), and thus removes instrumental and stellar rotation signals without affecting the phase curve.

\subsection{Posterior Sampling} \label{sec:sampling}

We construct the composite mean model and Gaussian process marginal likelihood in the \textsf{numpyro} framework with \textsf{JAX} \citep{jax2018github, numpyro}. The \textsf{numpyro} inference framework allows Python code to be compiled for efficient, parallel, Monte Carlo posterior sampling. \textsf{numpyro} provides support for several samplers, including standard Metropolis-Hastings and the No U-Turn Sampler (NUTS), which is a variation of the gradient-based Hamiltonian Monte Carlo (HMC) technique \citep[for a review of HMC, see][]{Betancourt2017}. We choose the No U-Turn Sampler for this work because in comparison with Metropolis-Hastings, for example, it often produces more effective samples from the posterior distribution in the less time.\footnote{In general, \textsf{JAX} can accelerate linear algebra computations on GPUs, and phase curve calculations can benefit from this acceleration. However, the calculation of the matrix inversion in the Gaussian process regression, which we describe in Section~\ref{sec:gp}, is memory-intensive and makes likelihood calculations quite expensive on the GPU. For these reasons, we run our \textsf{JAX} inference framework on the CPU.}

For each exoplanet, we construct phase curve models with the homogeneous and inhomogeneous reflected light parameterizations, and draw posterior samples with each model until the Gelman-Rubin statistic $\hat{r} < 1.01$ \citep{Gelman1992,Vehtari2019}.

\subsection{Injection-recovery test} \label{sec:inject}

\begin{figure*}
    \centering
    \includegraphics[width=\textwidth]{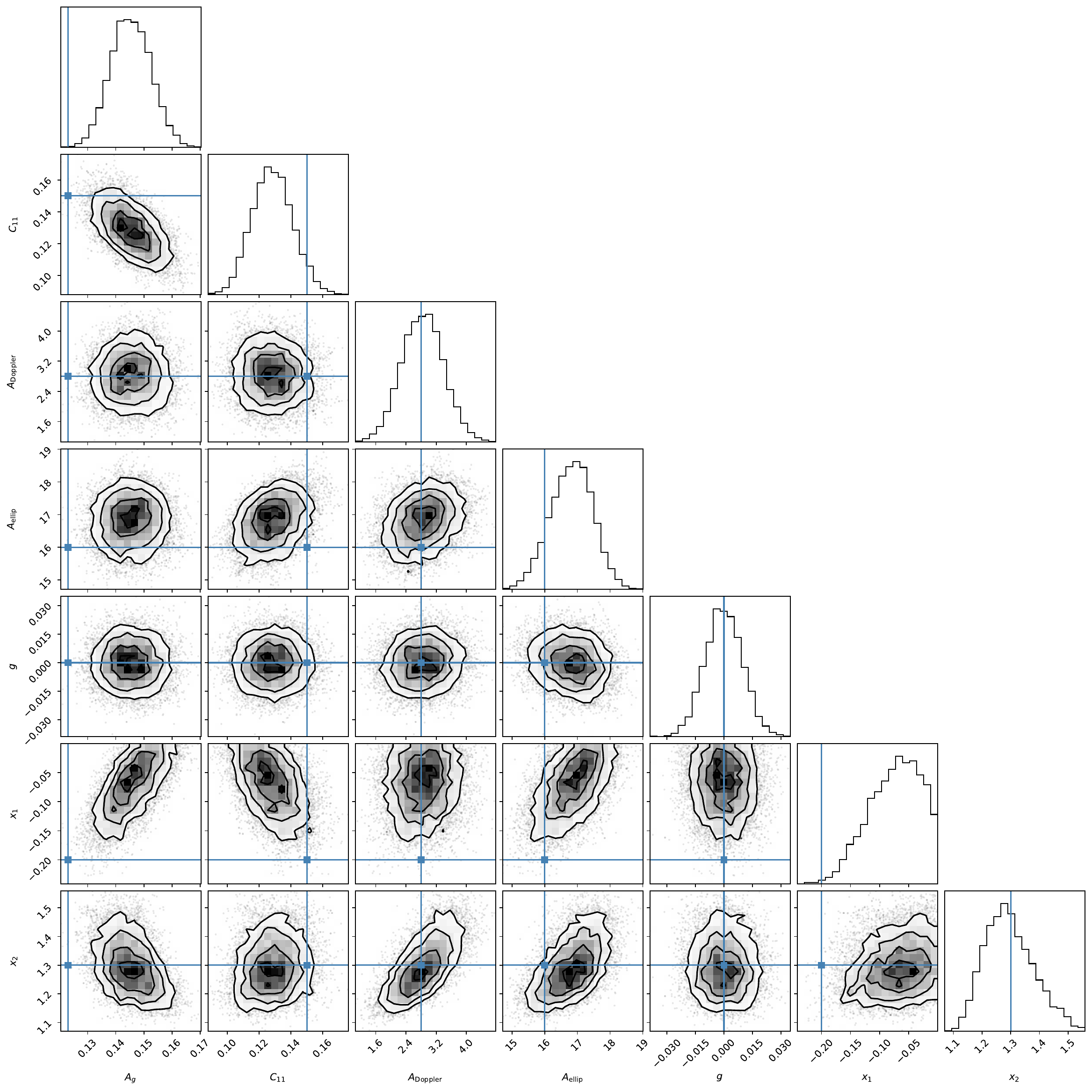}
    \caption{Posterior correlation plot demonstrating an injection-recovery exercise with a synthetic inhomogeneous phase curve signal injected into the real \kepler light curve of KIC 8424992. The blue lines indicate the values chosen for each of the injected phase curve parameters. The most challenging inference in this analysis is determining which fraction of the phase curve signal comes from reflection or thermal emission, since higher geometric albedos and hotter daysides both increase the planetary flux. This produces the anti-correlation between $C_{11}$ and $A_g$, since a cooler dayside can produce a similar phase curve to a more reflective atmosphere. The model produces accurate inferences despite this degeneracy by fitting for complementary parameters that define the shape of the phase curve at {\it all} phases, such as $x_1,~x_2,$ and $g$.}
    \label{fig:injection}
\end{figure*}

To evaluate the accuracy and precision of the \kelp framework for retrieving planetary atmospheric properties, we inject a synthetic phase curve into the real \kepler photometry of KIC~8424992. This G2V solar analog is a reasonable target for phase curve injections, as it is similar to the mostly Sun-like stars in this sample, with $K_p = 10.3$~mag, without known planets, and the following asteroseismic stellar properties: mass $0.9 \mathrm{~M}_\odot$, radius $1.05 \mathrm{~R}_\odot$, and age $9.8$ Gyr \citep{SilvaAguirre2017}. 

To stress-test the framework, we inject a synthetic phase curve into the light curve of KIC~8424992 with significant albedo asymmetry, substantial thermal emission, and significant stellar artifacts. We inject an inhomogeneous reflected light phase curve similar in asymmetry to Kepler-7 b, and we give the planet thermal emission and orbital properties similar to HAT-P-7 b for significant thermal emission and stellar artifacts.

The posterior distributions for each of the stellar and planetary atmosphere parameters are shown in Figure~\ref{fig:injection}, along with the ``true'' (injected) values marked in blue. Most parameters are retrieved within $2\sigma$ of their input values. The least accurate measurements are the geometric albedo $A_g$ ($3.1\sigma$), the start longitude of the darker region $x_1$ ($2.8\sigma$), and the spherical harmonic power $C_{11}$ ($1.7\sigma$), which are all degenerate with one other in the presence of ellipsoidal variations and Doppler beaming. These marginal inconsistencies arising from parameter degeneracies suggest that geometric albedos in this framework may have larger uncertainties than the posterior distributions suggest, up to a factor of a few.

\section{Results} \label{sec:results}

\begin{figure*}
    \centering
    \includegraphics[width=0.75\textwidth]{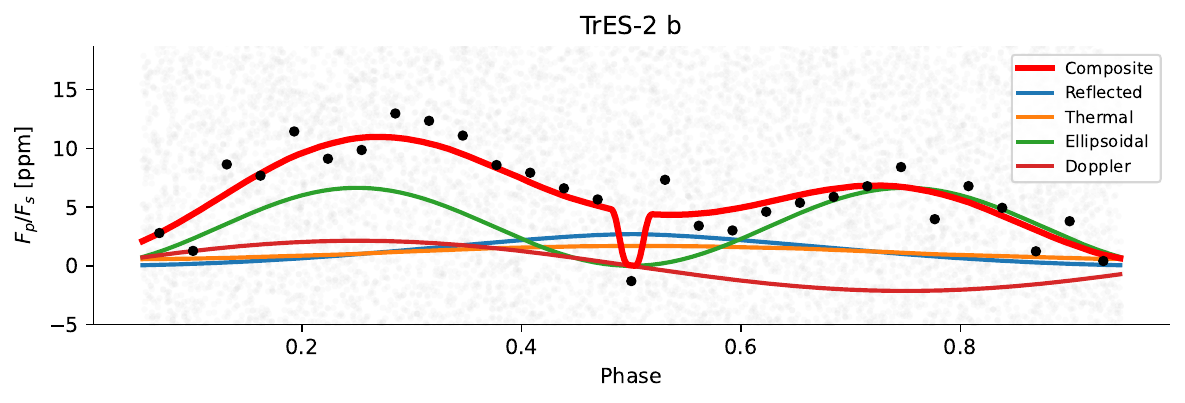}
    \includegraphics[width=0.75\textwidth]{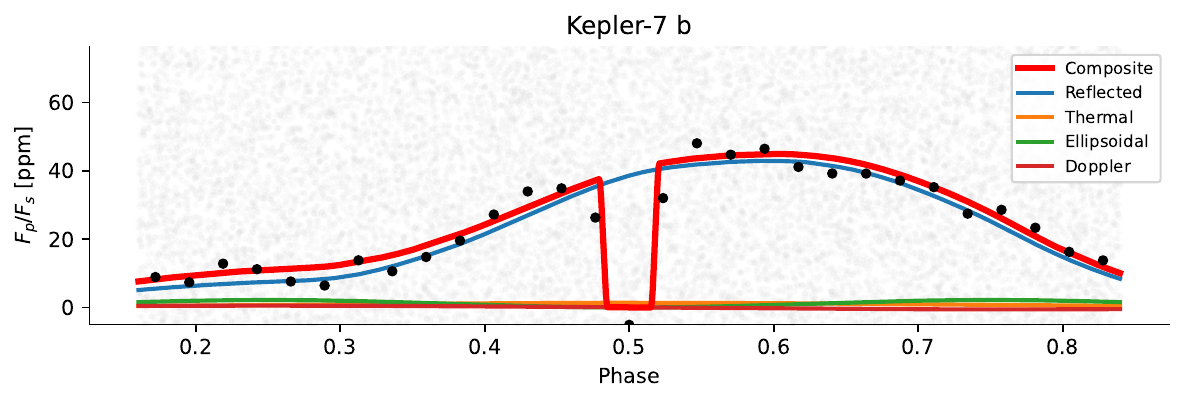}
    \includegraphics[width=0.75\textwidth]{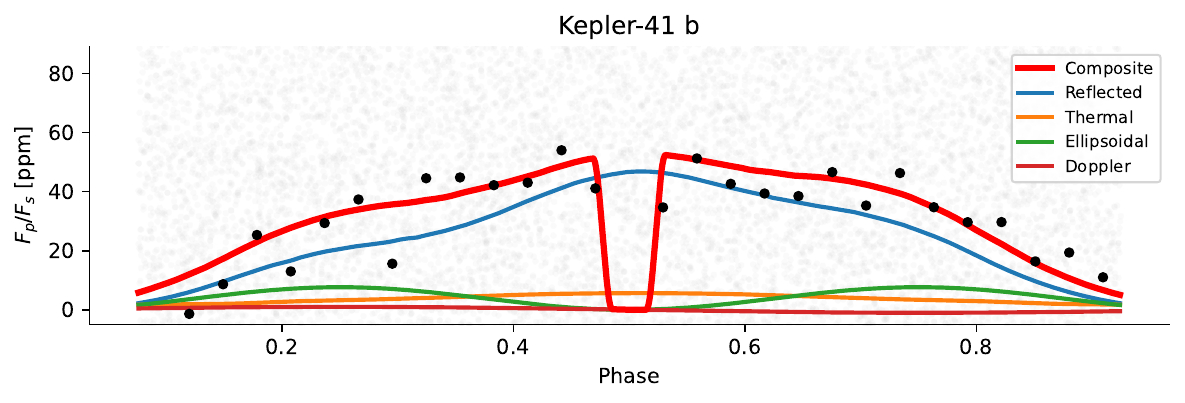}
    \includegraphics[width=0.75\textwidth]{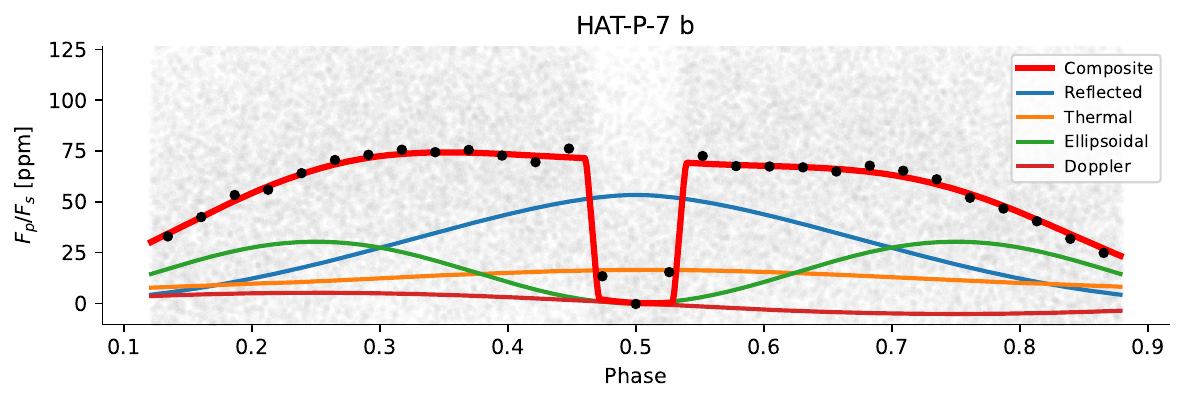}
    \includegraphics[width=0.75\textwidth]{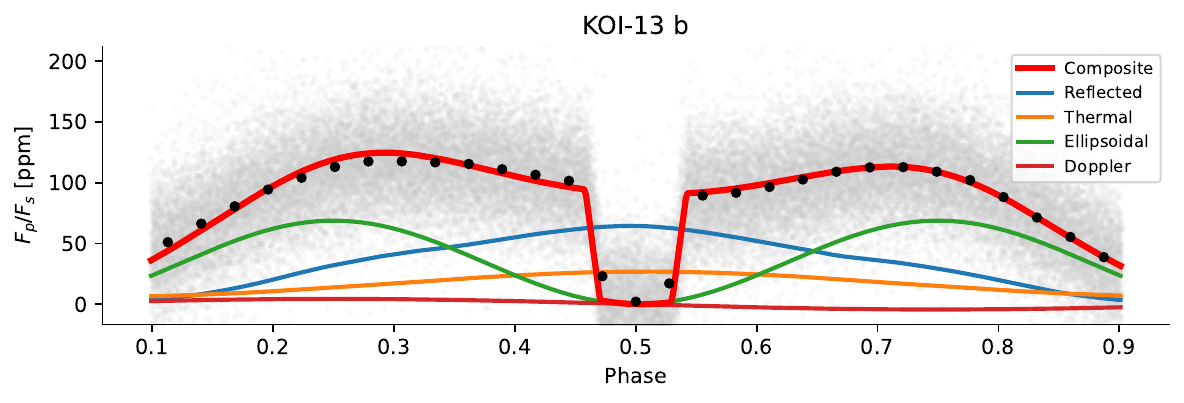}
    \caption{\kepler phase curves of five hot Jupiters after CBV detrending and removing the Gaussian process which captures stellar and instrumental artifacts. The four components of the phase curve are: reflected light (blue), thermal emission (orange), ellipsoidal variations (green), and Doppler beaming (dark red), along with the composite model in thick red. The unbinned 30-minute \kepler time series is shown in gray circles and is used in the fit, and median-binned photometry is shown with black circles for ease of visualization.}
    \label{fig:kepler-pcs}
\end{figure*}

\begin{figure*}
    \centering
    \includegraphics[width=\textwidth]{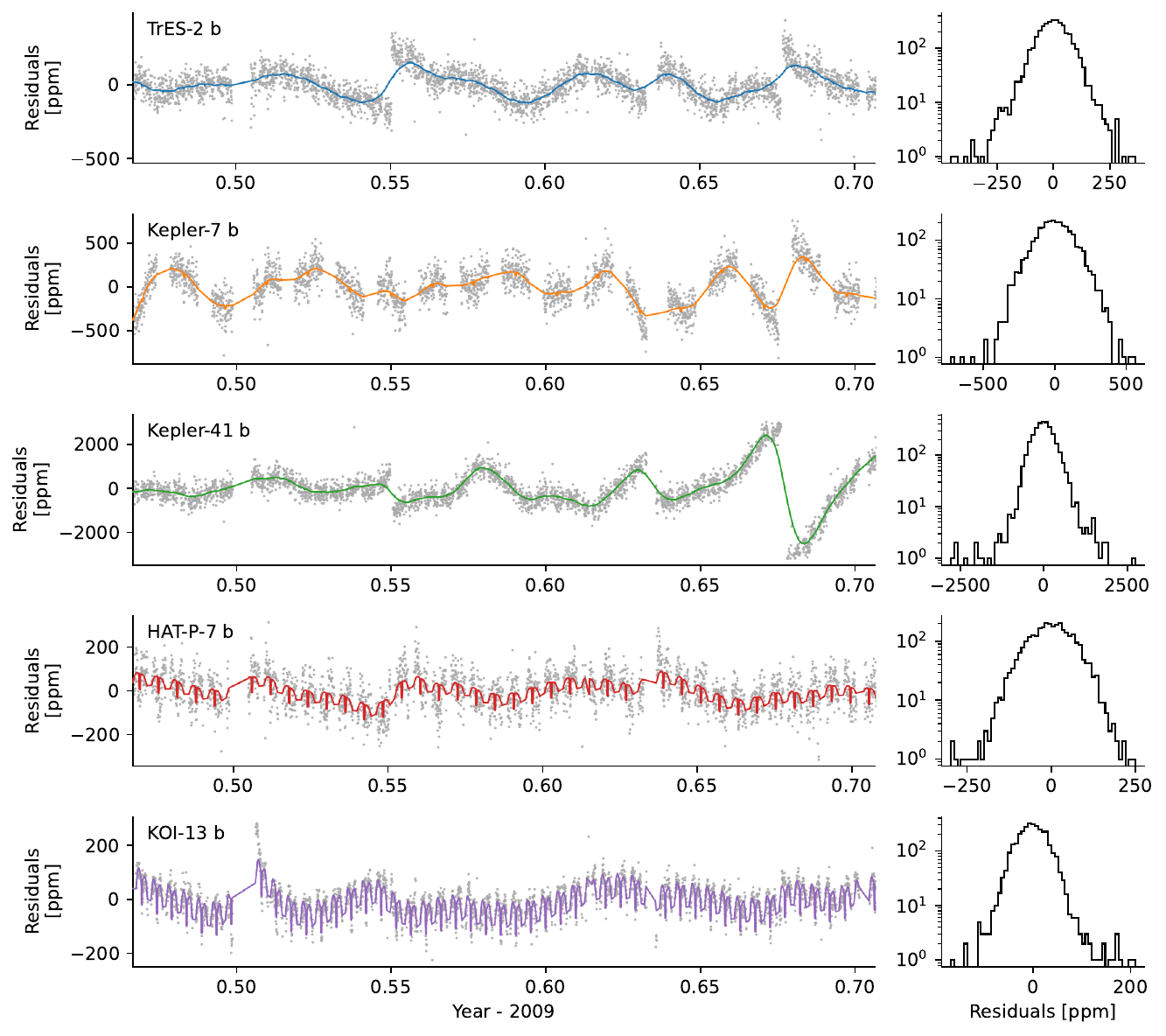}
    \caption{\kepler SAP fluxes (gray points) after CBV corrections are applied, in relative flux units of ppm, for Quarter 2 (roughly 6\% of each dataset). The composite fit (red curve) contains components including planetary phase curve (reflected light and thermal emission), strictly periodic stellar artifacts (ellipsoidal variations, Doppler beaming), and a Gaussian process for stochastic stellar artifacts such as magnetic activity (via a Gaussian process). The fit to the light curves on the left produces residuals which are shown in the histogram on the right.}
    \label{fig:residuals}
\end{figure*}

The maximum-likelihood models for all targets are shown in Figure~\ref{fig:kepler-pcs}, and the reflected light and thermal emission fitting parameters are enumerated in Table~\ref{tab:params}. The four phase curve components are plotted both separately and combined. The silver points are the raw \kepler time series after applying the CBV and GP corrections. No binning in time was applied to these long-cadence (30 minute) observations. 

The time series after CBV correction but before phase curves and GPs are removed is shown in Figure~\ref{fig:residuals}, in the time domain. Each light curve on the left shows one year of \kepler observations, and their corresponding residual histogram is plotted in the matching color on the right. With the exceptions of infrequent outliers, the log-histogram on the right is approximately quadratic, which indicates nearly-Gaussian noise.

\subsection{Leave-One-Out Cross-Validation} \label{sec:loo}

After we estimate posterior distributions for the phase curve parameters, we use cross validation and stacking to select between the homogeneous and inhomogeneous reflected light models. Two information criteria often used are Leave-One-Out Cross-Validation (LOO-CV) and the Widely Applicable (or Watanabe-Akaike) Information Criterion (WAIC). Following the recommendations in \citet{Vehtari2015b}, we choose Pareto-Smoothed Importance Sampling (PSIS) LOO-CV for this analysis \citep{Vehtari2015a}. Since we construct our likelihood with a Gaussian process, we compute the LOO for a non-factorized model with the efficient algorithm of \citet{Burkner2018}. We briefly summarize this procedure below.

PSIS-LOO-CV can be constructed as an iterative procedure where the $i$\textsuperscript{th} observation $y_i$ is left out, denoted with the negative subscript $y_{-i}$ to represent all observations except the $i$\textsuperscript{th}. The log-likelihood of the remaining observations is recomputed, repeating this calculation for all observations. The sum of these log-likelihoods gives the expected log pointwise predictive density, $\mathrm{elpd}_\mathrm{LOO}$. When the likelihood we are evaluating is the marginal likelihood of a Gaussian process, the model can not be trivially factorized, since the $\mathrm{elpd}_\mathrm{LOO}$ will depend on the draws for the GP hyperparameters. 

\citet{Burkner2018} show that for invertible covariance matrix $C$, one can define two quantities,
\begin{eqnarray}
   g_i &= \left[C^{-1} y\right]_i, \nonumber \\
  \bar{c}_{ii} &= \left[C^{-1}\right]_{ii}, \nonumber
\end{eqnarray}
where the $ii$ subscripts denote diagonal elements, such that the log predictive density becomes
\begin{eqnarray}
  \log p(y_i \,|\, y_{-i},\theta)
  = - \frac{1}{2}\log(2\pi)
  + \frac{1}{2}\log \bar{c}_{ii}
  - \frac{1}{2}\frac{g_i^2}{\bar{c}_{ii}}. \label{eqn:burkner}
\end{eqnarray}
Since this procedure involves inverting the full covariance matrix, it is expensive to compute, incurring computation times comparable to the integration for the posterior samples for these models. We perform this step ``offline'', after posterior sampling is complete. To further accelerate LOO-CV computation times, we make another simplifying assumption. We assume that the log pointwise predictive density can be computed on local subsets of the observations and then concatenated to produce an approximate global $\mathrm{lpd}_\mathrm{LOO}$. This assumption is based on the finite autocorrelation timescale of the Mat\'ern kernel. The observed fluxes are uncorrelated on timescales of several times $\rho_\mathrm{GP}$, and therefore the influence of widely-separated observations on the marginal likelihood should be negligible. 

In practice, we choose each \kepler Quarter -- three months of elapsed observing time -- to be the local subset of observations on which to compute each $\mathrm{lpd}_\mathrm{LOO}$. The size of the covariance matrix that must be inverted in Equation~\ref{eqn:burkner} would be the square of the full \kepler time series, of size $\approx44{,}000^2$ elements. Using the proposed subset LOO calculation reduces the size of the covariance matrix by a factor of $256$ to a more computationally inexpensive $\approx 2{,}700^2$ elements. This smaller matrix can be inverted on a desktop computer without special memory requirements. 

The batched, subset estimate of the LOO will be least accurate for observations at the beginning and end of \kepler quarters, on the ends of the batches. At these times, we will not propagate information into the marginal likelihood from autocorrelated astrophysical signals such as stellar rotation that are coherent across multiple \kepler quarters. Most stars considered in this work are solar-type stars and rarely have activity signals that are coherent for more than one month \citep{Giles2017, Morris2019c} -- see Figure~\ref{fig:residuals}. Thus we posit that the subset LOO estimates may be sufficient for model selection in this case. 

\subsection{Bayesian model selection} \label{sec:modelselection}

When several models are available to describe the observations, one may discard disfavored models by means of Bayesian model selection. \citet{Yao2018} recommend the stacking after comparing three types of Bayesian model averaging (BMA) with the model stacking technique. Both stacking and BMA methods yield values called the ``model weights.'' The model weights must sum to unity, and a model with weight close to zero can be rejected in favor of models with weights close to unity. Intermediate weights indicate no strong preference for either model. 

We compute the model weights using a fork of the \textsf{arviz.compare} function. To confirm that stacking works within our framework, we verified that the PSIS-LOO-CV and stacking techniques indicate a clear preference for the inhomogeneous model when applied to the injection/recovery tests in Section~\ref{sec:inject}, producing an inhomogeneous model weight close to unity for a phase curve with significant, prescribed inhomogeneity.

\begin{table*}
\centering
\begin{tabular}{lccccc}
{\bf Parameter} & {\bf TrES-2 b} & {\bf Kepler-7 b} & {\bf Kepler-41 b} & {\bf HAT-P-7 b} & {\bf KOI-13 b} \\ \hline \\
\multicolumn{6}{l}{\bf Model comparison results} \\\\
Homogeneity & Hom. & Inhom. & Inhom. & Hom. & Inhom. \\\\
Phase Func. & C-S. & C-S. & Iso. & Iso. & C-S. \\\\
Weight Inh. & 0.00 & 0.90 & 1.00 & 0.28 & 0.72 \\\\
Weight C-S & 1.00 & 1.00 & 0.00 & 0.00 & 0.77 \\\\ \hline \\
\multicolumn{6}{l}{\bf Posterior distributions} \\\\
$\omega$ & ${0.06}_{-0.03}^{+0.02}$ & ${0.99}_{-0.01}^{+0.00}$ & ${0.92}_{-0.06}^{+0.04}$ & ${0.40}_{-0.07}^{+0.07}$ & ${0.98}_{-0.02}^{+0.01}$ \\\\
$A_g$ & ${0.01}_{-0.01}^{+0.00}$ & ${0.26}_{-0.01}^{+0.01}$ & ${0.13}_{-0.02}^{+0.01}$ & ${0.09}_{-0.02}^{+0.02}$ & ${0.32}_{-0.03}^{+0.03}$ \\\\
$x_1$ [deg] & -- & ${-8.23}_{-3.19}^{+3.55}$ & ${-29.44}_{-5.51}^{+6.95}$ & -- & ${-21.03}_{-3.89}^{+3.44}$ \\\\
$x_2$ [deg] & -- & ${56.76}_{-4.10}^{+5.24}$ & ${38.80}_{-4.34}^{+4.50}$ & -- & ${12.77}_{-6.81}^{+5.72}$ \\\\
$C_{11}$ & ${0.05}_{-0.03}^{+0.05}$ & ${0.06}_{-0.04}^{+0.07}$ & ${0.07}_{-0.05}^{+0.06}$ & ${0.14}_{-0.02}^{+0.02}$ & ${0.10}_{-0.02}^{+0.02}$ \\\\ 
$f$ & ${0.71}_{-0.01}^{+0.01}$ & ${0.71}_{-0.01}^{+0.01}$ & ${0.71}_{-0.01}^{+0.01}$ & ${0.70}_{-0.01}^{+0.01}$ & ${0.70}_{-0.01}^{+0.01}$ \\\\ \hline \\
\multicolumn{6}{l}{\bf Derived quantities} \\\\
Reflected light phase offset [deg] & -- & ${34.42}_{-4.14}^{+2.38}$ & ${2.00}_{-1.21}^{+1.40}$ & -- & ${-1.24}_{-0.43}^{+0.32}$ \\\\
Reflected light amplitude [ppm] & ${2.21}_{-0.98}^{+1.24}$ & ${11.71}_{-0.59}^{+0.61}$ & ${5.20}_{-0.85}^{+0.62}$ & ${47.76}_{-9.02}^{+11.63}$ & ${14.21}_{-1.67}^{+1.25}$ \\\\
Thermal emission amplitude [ppm] & ${1.61}_{-0.51}^{+1.05}$ & ${1.19}_{-0.44}^{+1.24}$ & ${6.44}_{-2.89}^{+6.35}$ & ${20.27}_{-3.72}^{+5.17}$ & ${51.14}_{-8.50}^{+8.48}$ \\\\
$T_d$ [K] & ${1653}_{-50}^{+70}$ & ${1730}_{-67}^{+108}$ & ${1925}_{-102}^{+127}$ & ${2408}_{-55}^{+62}$ & ${2812}_{-64}^{+56}$ \\\\
$T_n$ [K] & ${1510}_{-69}^{+49}$ & ${1531}_{-108}^{+65}$ & ${1652}_{-126}^{+100}$ & ${2138}_{-63}^{+53}$ & ${2289}_{-55}^{+64}$ \\\\
\end{tabular}
\caption{Reflected light and thermal emission phase curve parameters inferred from \kepler observations. For prior distributions, see Table~\ref{tab:priors}. The first section gives results from model comparison with leave-one-out cross-validation (LOO-CV) and model stacking. Models are marked "Hom." when LOO-CV prefers the homogeneous reflected light model, "Inhom." when LOO-CV prefers the inhomogeneous model. The scattering phase function is included when preferred by LOO-CV with the same criteria. The posterior distributions for the reflected light and thermal emission parameters are in the second section, excluding $g$ which was consistent with the prior for all planets. The derived quantities include the phase offset of the reflected light phase curve, the amplitudes of the reflected light and thermal emission components, and the dayside and nightside temperatures. See Section~\ref{sec:inject} for discussion on the uncertainties derived from these degenerate posterior distribution. \label{tab:params}}
\end{table*}

\begin{figure*}
    \centering
    \includegraphics[width=\textwidth]{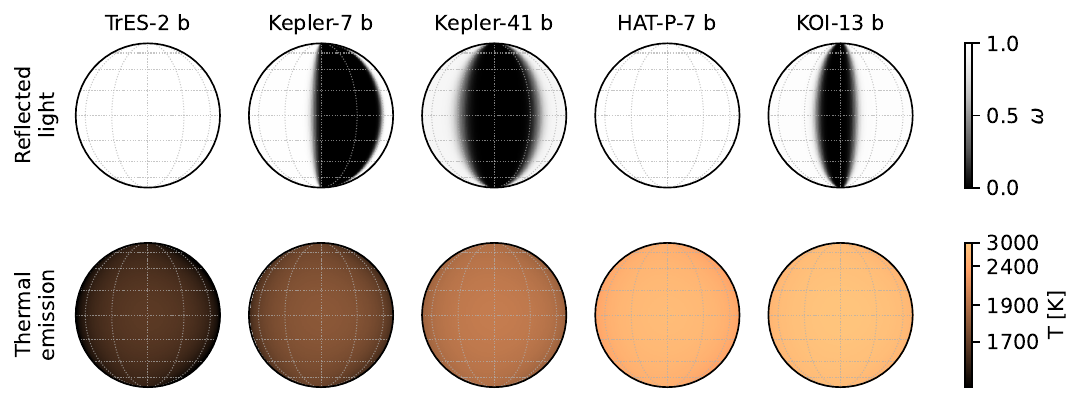}
    \caption{Inferred reflectance and temperature maps for each planet, where the center of the map corresponds to the sub-stellar point. The shading in the upper row corresponds to the single scattering albedo $\omega$ of the reflective region on the dayside. Cross validation prefers inhomogeneous models for each planet, so each dayside features a dark sub-stellar region bounded by sub-observer longitudes $x_1$ and $x_2$. The thermal emission maps in the second row are not uniform, but appear nearly uniform in this figure because the colormap spans the full range of dayside temperatures for all five planets.}
    \label{fig:maps}
\end{figure*}

\subsection{Homogeneity}

\begin{figure}
    \centering
    \includegraphics[width=\columnwidth]{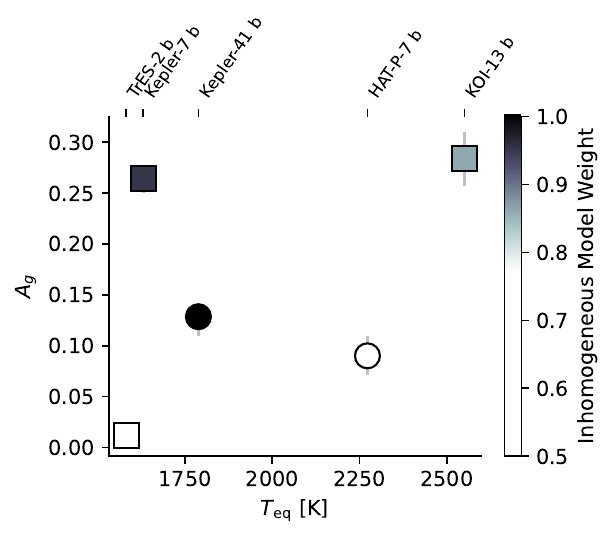}
    \caption{Geometric albedos and equilibrium temperatures for the five \kepler planets, colored by the inhomogeneous model weight. Model comparisons with weights approaching unity confidently prefer the inhomogeneous model over the homogenous model (see Section~\ref{sec:modelselection}), so darker points are more likely inhomogeneous reflectors. The marker shape corresponds to the preferred scattering phase function: squares are Cornette-Shanks single scattering plus isotropic multiple scattering, and circles are isotropic single plus isotropic multiple scattering. Results are also enumerated in Table~\ref{tab:params}.}
    \label{fig:homogeneity}
\end{figure}

Using the above LOO-CV and stacking techniques (see Sect.~\ref{sec:loo}), we can efficiently compute the posterior predictive density for the full model including the GP, and determine which model best describes the observations. We compare models which assume the planet is an inhomogeneous reflector and models assuming homogeneous reflectors for each planet. The comparison is plotted in Figure~\ref{fig:homogeneity} which shows darker blue points for planets more likely to be inhomogeneous reflectors. The results are also enumerated in Table~\ref{tab:params}.

The cross-validation only conclusively prefers the homogeneous phase curve model for TrES-2 b. The homogeneity selection is least confident about HAT-P-7 b and KOI-13 b (Table~\ref{tab:params}), since weights near 0.5 indicate no preference for either of the two models. We expect that the two least-certain homogeneity inferences occur for the two closest-in and hottest planets because they produce most thermal emission, and the largest ellipsoidal variations and Doppler beaming, all of which exacerbate degeneracies in the phase curve shape.

Theory also suggests less confident expectations for clouds on HAT-P-7 b and KOI-13 b. With dayside temperatures $T_d > 2500$ K, their atmospheres are likely too hot for condensates to persist, so we might expect them to be cloud-free, homogeneous reflectors (see for example \citealt{Marley2013cctp.book..367M}).

\begin{figure}
    \centering
    \includegraphics[width=\columnwidth]{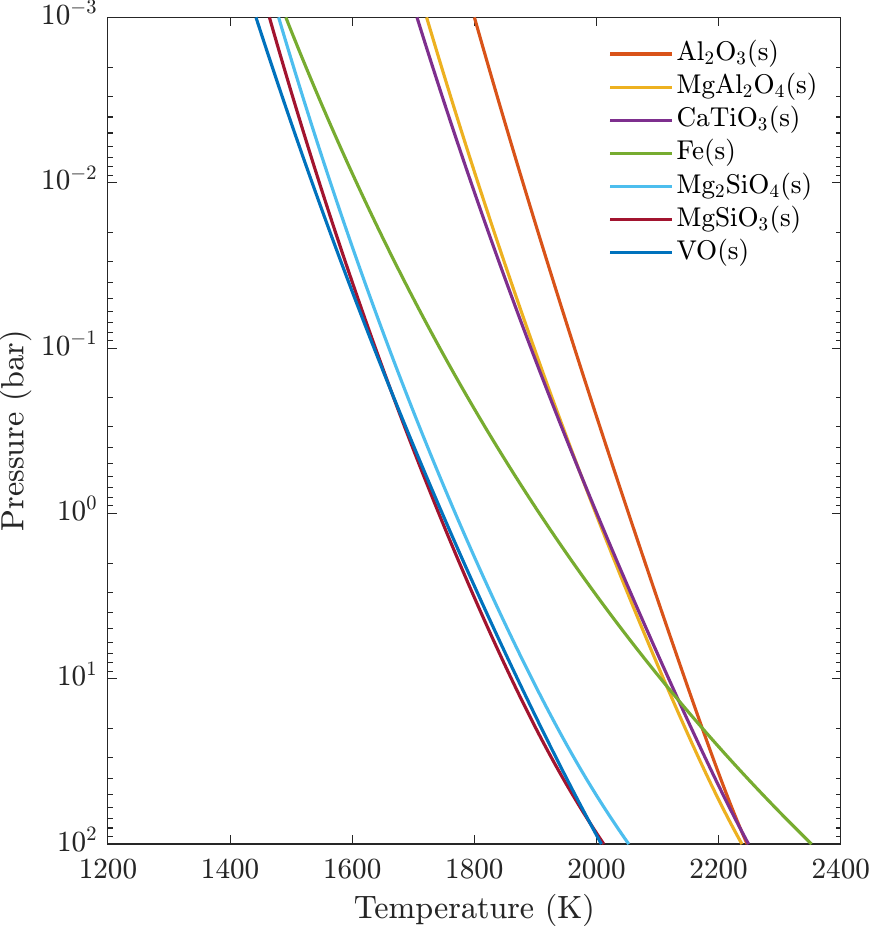}
    \caption{Stability curves of selected high-temperature condensates for an atmosphere with the metallicity of Kepler-41 b $[\mathrm{M/H}]=0.38$ \citep{Bonomo2015}. The calculations are assuming chemical equilibrium in the gas phase.}
    \label{fig:hat7-tp}
\end{figure}

The inhomogeneity classification for HAT-P-7 b presents an interesting test case. With an equilibrium temperature of $T_\mathrm{eq} \approx 2300$ K, HAT-P-7 b resides close to the temperature above which we do not expect condensates to form. Dayside emission spectroscopy of HAT-P-7 b, like that of \citet{Mansfield2018}, can be used to infer the temperature-pressure (p-T) structure for the atmosphere, which may be greater than this condensation limit \citep[see for example Figure~8 of][]{Christiansen2010}, but we emphasize here that these p-T profiles are typically a 1D dayside average, and do not account for the local surface temperature variations that we know must exist on real hot Jupiters. While the observations in this work indicate inhomogeneity, theoretical expectations for inhomogeneity are unclear this planet.

Kepler-7 b is perhaps one of the better known reflected light phase curves in the literature due to its high albedo, cool equilibrium temperature and strong reflected light asymmetry \citep{Demory2013, Esteves2015, Hu2015, Angerhausen2015, Heng2021}. As the target with the most clearly asymmetric phase curve, it is ideal for confirming that the model selection framework is performing correctly, and indeed the framework yields a strong preference for the inhomogeneous model. This may also be an important test of the GP flexibility -- if the GP is too flexible, it could partially or entirely absorb the asymmetry in the phase curve of Kepler-7 b. It appears that the GP works as intended, and does not interfere with the short-period phase curve oscillations. 

The confident detection of inhomogeneity for Kepler-41 b is in agreement with \citet{Shporer2015}, who pointed out that such an inference could be drawn from earlier works on the system \citep{Santerne2011,Quintana2013}. We comment on the results for Kepler-41 b in more detail in Section~\ref{sec:scatterers}.

TrES-2 b is remarkable for its very low albedo \citep{Kipping2011}. As the coolest planet in the sample, it should also have the least contamination from thermal emission in the \kepler bandpass. The $3\sigma$ upper limit on the geometric albedo is $2\%$. The LOO-CV analysis indicates a strong preference for inhomogeneity.

KOI-13 A is unique among the host stars for being a hierarchical triple, with the star the A component hosting the transiting planet \citep{Shporer2014}. We adopt the dilution due to the other components of the triple system from the analysis of \citet{Shporer2014}, described in Section~\ref{sec:contamination}. 

\subsection{Scattering phase functions} \label{sec:scattering-phase-functions}

\citet{Heng2021} showed that the shape of the phase curve encodes information about the scattering phase function of the atmosphere. We examine here whether the phase curves contain sufficient information to identify the scattering phase function in each reflected light phase curve. Following \citet{Hapke1981}, \citet{Heng2021} assumes that multiple scattering is always isotropic, while the single-scattering component could be described by any phase function. We thus probe two independent questions in this subsection: (1) is isotropic multiple scattering a good assumption, and (2) do the observations prefer Rayleigh or \citet{Cornette1992} single scattering over isotropic?

\subsubsection{Theoretical justification for isotropic multiple scattering} \label{sec:theory_isotropic}

To test the applicability of the isotropic multiple scattering assumption made by \citet{Heng2021}, we perform additional numerical simulations for Rayleigh and isotropic scattering phase functions. 
For the numerical calculations we use the discrete ordinate radiative transfer model \textsc{C-Disort} \citep{Hamre2013AIPC.1531..923H}, which is the C-version of the well-established \textsc{Disort} model \citep{Stamnes1988ApOpt}.

\textsc{Disort} provides the exact solution to the plane-parallel radiative transfer equation, taking into account thermal emission, surface scattering, and illumination by a beam source at the top of the atmosphere. It supports general scattering phase functions, described by the usual expansion into a Legendre series. Given a number of computational streams (ordinates), \textsc{Disort} provides the full, angular-resolved intensity field $I(\mu, \phi)$, where $\mu$ is the cosine of the polar and $\phi$ the azimuth angle. 

In order to simulate a semi-infinite atmosphere \citep{Chandrasekhar1960ratr.book.....C} as done in the calculations presented in \citet{Heng2021}, we set the optical depth at the bottom of the atmosphere to $\tau = 10^6$. The surface albedo is set to zero, but in principle plays no role for the resulting outgoing radiation due to the very high optical depth assumed at the lower boundary. Since we are only interested in the reflected light phase curve, we turn off the thermal emission term within \textsc{Disort} and set the incident stellar beam as an exterior source. 

\textsc{Disort} calculates the radiation field in local coordinates rather than in the observer's coordinate system as described in \citet{Heng2021}. We therefore use Equation (13) from \citet{Heng2021}, which is originally from \citet{Sobolev1975lpsa.book.....S}, to convert the observer-centric phase angle $\alpha$, longitude $\Phi$, and latitude $\Theta$ to the local coordinates for a given atmospheric column. This applies in particular to the local zenith angle $\mu_*$ of the stellar beam. From the \textsc{Disort} calculations we then extract the intensity at the polar angle $\mu(\Phi, \Theta)$ and azimuth angle $\phi(\Phi, \Theta, \alpha)$ scattered into the direction of the observer. The corresponding angles are again obtained by using Equation (13) from \citet{Heng2021}. 

The radiative transfer is calculated for a range of values for longitude, latitude, and phase angle. The resulting intensities reflected towards the observer as a function of $\alpha$, $\Phi$, and $\Theta$ are transformed into the reflection coefficient $\rho$ and then numerically integrated over $\Phi$ and $\Theta$ to yield the Sobolev fluxes $F(\alpha)$, following Equation (5) from \citet{Heng2021}, which is originally from \citet{Sobolev1975lpsa.book.....S}. Finally, these fluxes are converted into the integral phase function
\begin{equation}
  \Psi(\alpha) = \frac{F\left(\alpha \right)}{F\left( \alpha = 0 \right)} \ .
\end{equation}

\begin{figure}
    \centering
    \includegraphics[width=\columnwidth]{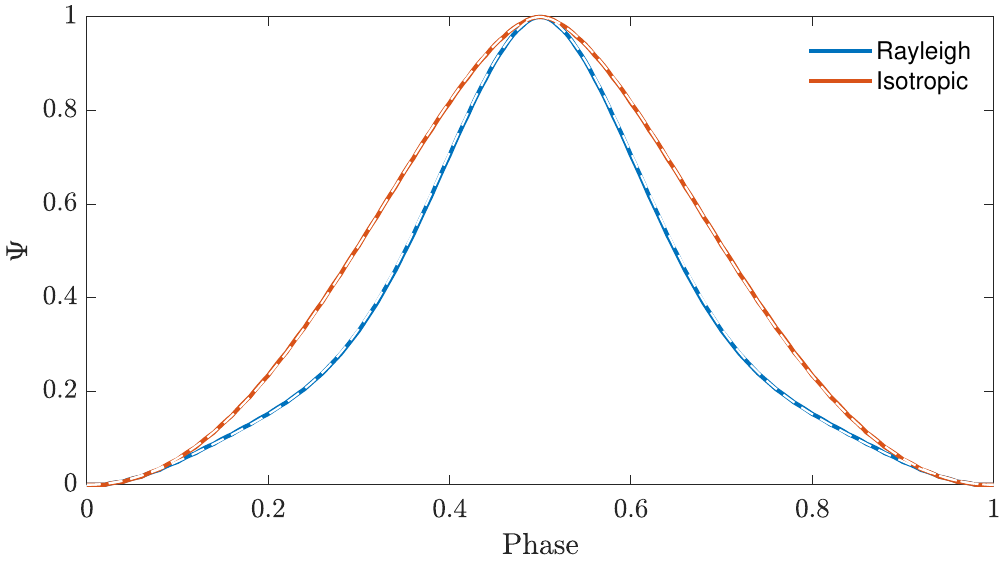}
    \caption{Comparing two simple scattering phase functions for a single-scattering albedo of $\omega = 0.5$ with both single and multiple scattering (solid curves) and with different single scattering phase functions but forcing isotropic multiple scattering (dashed, white curves), as in \citet{Heng2021}. The difference between ``full'' Rayleigh scattering and Rayleigh single plus isotropic multiple scattering is negligible.}
    \label{fig:phase-curve-comparison}
\end{figure}

In Figure~\ref{fig:phase-curve-comparison}, we compare the outcome for $\Psi$ of numerical calculations performed with \textsc{Disort} with the analytical phase curve description from \citet{Heng2021} for Rayleigh and isotropic scattering phase functions. These calculations are identical to the scenario presented in Figure 3(a) from \citet{Heng2021}, assuming a constant single scattering albedo of $\omega = 0.5$ throughout the atmosphere.

The analytic phase curves of \citet{Heng2021} are in excellent agreement with the numerical results. The assumption of isotropic multiple scattering is a good approximation for full-Rayleigh scattering. While the Rayleigh scattering phase function shows small deviations from an isotropic behavior, this  is only important for single scattering events. When averaged over many multiple scattering events in a semi-infinite atmosphere, however, the phase function tends towards a more isotropic behavior.

\subsubsection{Single scattering phase function}

For all planets, using the LOO-CV and stacking techniques detailed above, the Cornette-Shanks or isotropic single scattering phase functions are clearly preferred, with Bayesian stacking model weights close to unity. The preference for isotropic single scattering is rather surprising at first, since no known atmospheric constituent, molecules, or potential condensates are known to produce a truly isotropic scattering phase function. 

Molecules scatter radiation according to the Rayleigh scattering phase function. For condensates, assuming a spherical shape in the simplest case, Mie scattering is applicable \citep{Mie1908AnP...330..377M}. When particles are small compared to the wavelength, the Mie scattering phase function simplifies to the Rayleigh one \citep{Bohren1998asls.book.....B}. For example, a hypothetical homogeneously-reflecting planet may be uniformly cloud-free, and reflected light from cloudless giant planets should be dominated by Rayleigh scattering from H$_2$ and He, which we expect to scatter with a Rayleigh phase function.

However, we should expect that multiple scattering is significant for several of these objects, as they have non-zero values of the single scattering albedo $\omega$. As shown in the previous subsection, even pure Rayleigh single and Rayleigh multiple scattering will produce a phase curve similar to isotropic multiple scattering. The difference between Rayleigh and isotropic single scattering is maximal near quadrature but on the order of $10\%$ in amplitude. We expect such subtle features in the shape of the phase curve to be degenerate with the other phase curve features such as the ellipsoidal variations in the presence of photometric noise. As a result, the preference for or against isotropic scattering could be seen as a weak test of the cloudiness of an atmosphere, though the clear asymmetry in the phase curves of a planet like Kepler-7 b is a much less ambiguous test of inhomogeneity.

In \citet{Heng2021}, the Henyey-Greenstein (HG) scattering phase function was adopted for Kepler-7 b. For scattering asymmetry parameters $g \rightarrow 0$, the HG function approaches an isotropic scattering phase function. The prior on $g$ in \citet{Heng2021} therefore enforced that the scattering phase function was close to isotropic. This work uses a Cornette-Shanks phase function, so the strong prior $g \rightarrow 0$ enforces a nearly-Rayleigh phase function.

\subsection{Variability}

\begin{figure}
    \centering
    \includegraphics[width=\columnwidth]{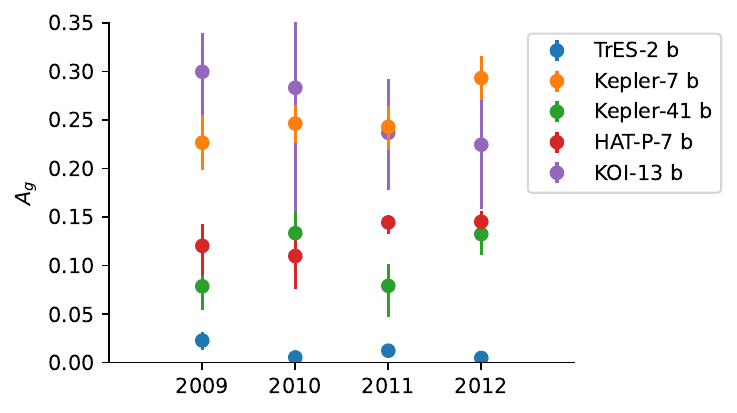}
    \caption{Geometric albedos for four \kepler targets, measured in four one-year segments of the \kepler light curves.} 
    \label{fig:yearly-A_g}
\end{figure}

The best-fit phase curves are in Figure~\ref{fig:kepler-pcs} (thick red curves), which are comprised of a combination of reflected light, thermal emission, ellipsoidal variations, and Doppler beaming (thin curves). All models are fit to the unbinned, 30-minute long cadence \kepler photometry with a Gaussian process in the time domain and CBV correction.

Figure~\ref{fig:residuals} shows the CBV-corrected SAP fluxes from the \kepler mission along with a sample model which has a Gaussian process with a mean model described by the four phase curve components: reflection, thermal emission, Doppler beaming and ellipsoidal variations. The light curves are separated into four segments and labeled with the year at the beginning of each year of observations. The residuals after the phase curve and Gaussian process model have been applied are plotted as a histogram on the right of Figure~\ref{fig:residuals}, showing consistent, Gaussian normal distributions.

The large-amplitude signals in the CBV-corrected SAP fluxes in Figure~\ref{fig:residuals} may have stellar origins. Kepler-41, a G2V host, shows a consistent autocorrelated behavior in the flux residuals with amplitude 2 ppt at all times during the \kepler observations. G0V star TrES-2 shows possible rotational signals in the first and fourth year of \kepler observations. HAT-P-7 (F8V) and KOI-13 A (A5-7V) show 250 ppm variations which may be rotational or instrumental. In all cases, the Gaussian process accounts for this autocorrelated signal on timescales of tens of days, without significantly affecting phase curve signals on one-day timescales.

The geometric albedos we infer for each planet for each year are plotted in Figure~\ref{fig:yearly-A_g} and listed in Table~\ref{tab:ag}. There are no significant departures from a constant geometric albedo for any of the four planets. 

\begin{table*}
\centering
\begin{tabular}{l|cccc}
 & \multicolumn{4}{c}{Geometric albedo}\\\\
Planet & Q1-4 & Q5-8 & Q9-12 & Q13-16 \\ \hline \\
TrES-2 b & $ 0.02 \pm  0.01$ & $ 0.01 \pm  0.00$ & $ 0.01 \pm  0.01$ & $ 0.01 \pm  0.00$ \\\\
Kepler-7 b & $ 0.23 \pm  0.03$ & $ 0.25 \pm  0.02$ & $ 0.24 \pm  0.02$ & $ 0.29 \pm  0.02$ \\\\
Kepler-41 b & $ 0.08 \pm  0.02$ & $ 0.13 \pm  0.03$ & $ 0.08 \pm  0.03$ & $ 0.13 \pm  0.02$ \\\\
HAT-P-7 b & $ 0.12 \pm  0.03$ & $ 0.11 \pm  0.03$ & $ 0.14 \pm  0.01$ & $ 0.14 \pm  0.02$ \\\\
KOI-13 b & $ 0.30 \pm  0.04$ & $ 0.24 \pm  0.11$ & $ 0.24 \pm  0.05$ & $ 0.22 \pm  0.06$ \\\\
\end{tabular}
\caption{Geometric albedos inferred independently for one-year intervals for five \kepler hot Jupiters.  \label{tab:ag}}
\end{table*}

\begin{figure}
    \centering
    \includegraphics[width=\columnwidth]{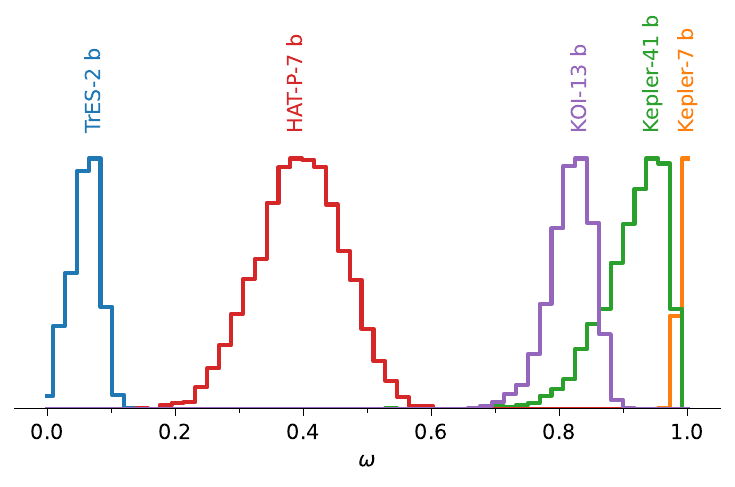}
    \caption{Posterior distributions for the single scattering albedos. For inhomogeneous atmospheres, $\omega$ is the single scattering albedo of the more reflective region towards the limbs, and for homogeneous atmospheres $\omega$ is the dayside hemisphere-averaged single scattering albedo. For comparison, the single scattering albedos of plausible condensates are shown in Figure~\ref{fig:lx-mie}. }
    \label{fig:omegas}
\end{figure}

\section{Interpretation} \label{sec:interpretation}

The techniques in this work advance our ability to characterize exoplanet atmospheres with photometry alone. In this section, we provide interpretation of the results presented in the previous section, assisted by condensation and Mie scattering calculations, to interpret what species might be responsible for inhomogeneous scattering. We also outline the observational biases introduced by the scattering parameters used in this work. 

\subsection{Identifying potential scattering species} \label{sec:scatterers}

\begin{figure*}
    \centering
    \includegraphics[width=\textwidth]{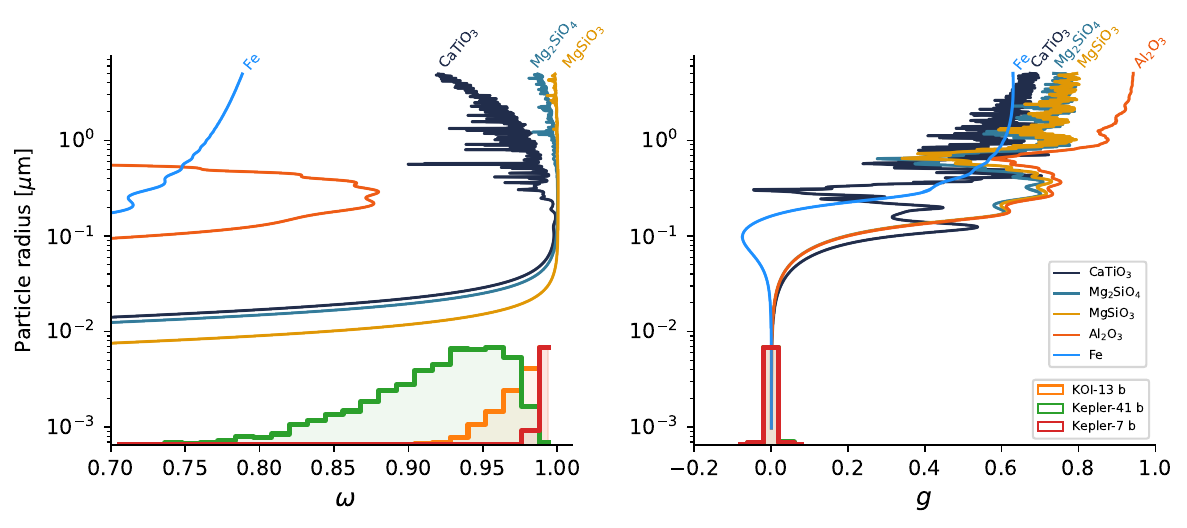}
    \caption{Predictions from Mie theory for the reflectance properties of several refractory species assuming spherical, monodisperse particles observed in the \kepler bandpass as a function of particle radius \citep[][curves above]{Kitzmann2018a}, for comparison with the posterior distributions for the reflectance parameters from \kepler observations (histograms). The measured single scattering albedos $\omega$ of each planet (also shown in Figure~\ref{fig:omegas}) have $0.7 \lesssim \omega \lesssim 1$, placing a lower limit on the particle radius $r \gtrsim 0.01~\mu\mathrm{m}$. The phase curves are consistent with scattering asymmetry parameters $g < 0.2$, though posterior distributions are so narrow as a result of the prior distribution. Fits consistent with near-zero $g$ correspond to upper limits on particle radii of $r \lesssim 0.1 \mu\mathrm{m}$. The three species consistent with the phase curve observations for Kepler-7 b and Kepler-41 b are CaTiO$_3$ (perovskite), Mg$_2$SiO$_4$ (forsterite), and MgSiO$_3$ (enstatite), given that they have $\omega > 0.9$ and near-zero $g$. We compare the condensation temperatures of each of these species with the local temperatures at the longitudes of the less reflective region in Figure~\ref{fig:condensation-longitudes}.}
    \label{fig:lx-mie}
\end{figure*}

Theoretical expectations and the observational evidence outlined above point to inhomogeneous albedo distributions for all planets. For planets in this temperature range, we expect that differential condensation with longitude will produce inhomogeneous albedo distributions.  At the cooler end the sample where $T_\mathrm{eq} < 2000$ K, there are several refractory species that one might expect to condense out of regions in the dayside atmospheres \citep{Marley2013cctp.book..367M}. 

Figure~\ref{fig:hat7-tp} shows stability curves for an atmosphere with $[\mathrm{M/H}]=0.38$. The curves indicate temperatures and pressures where condensation may occur, and are calculated with the chemical equilibrium model \textsc{FastChem}\footnote{\url{https://github.com/exoclime/FastChem}} \citep{Stock2018MNRAS.479..865S, Stock2022, Kitzmann2023arXiv230902337K}, using the JANAF tables \citep{Cha98} to derive the mass action constants for selected condensates as described by \citet{Kitzmann2023arXiv230902337K}. Some of the most stable high-temperature condensate species across all pressures are Al$_2$O$_3$ (corundum), CaTiO$_3$ (perovskite), MgAl$_2$O$_4$ (spinel), Mg$_2$SiO$_4$ (forsterite), MgSiO$_3$ (enstatite), and Fe(s) (see e.g.  \citealt{Fortney2005} or \citealt{Gail2013pccd.book.....G}). Each species condenses at temperatures $ T \lesssim 2000$ K at 1~bar, so they are all possible condensates for Kepler-7 b, Kepler-41 b and TreS-2 b. We can further narrow the list of possible condensates and their properties with two methods outlined below.

The first method is to predict the possible values of $\omega$ and $g$ for each species using Mie theory, and compare them with the observations. Given the indices of refraction for a given species, the wavelength of observations, and the particle size, one can directly compute the expected value for $\omega$ and $g$ using \textsf{LX-MIE} \citep{Kitzmann2018a}. The predictions are shown in Figure~\ref{fig:lx-mie} for five refractory species which may condense in these exoplanet atmospheres, as a function of particle size. 

Figure~\ref{fig:lx-mie} allows us to roughly constrain the sizes of condensate particles in the exoplanet atmospheres. All inhomogeneous planet have single scattering albedos $\omega \gtrsim 0.9$. This provides a lower limit on the particle size since all compatible albedos occur for these scattering species when their particle radii $r \gtrsim 0.01~\mu$m. Similarly, we can place a rough upper limit on the particle size of condensates, since each phase curve can be reproduced with small $g$, excluding large particle sizes $r \gtrsim 0.2~\mu$m. The Mie calculations therefore weakly constrain the cloud particle radius between $10^{-2} - 10^{-1}~\mu$m. 

The Mie calculations disfavor Al$_2$O$_3$ (corundum) and Fe (iron) as the condensates responsible for reflection from Kepler-41 b and Kepler-7 b. These atmospheres have single scattering albedos $\omega \gtrsim 0.9$, which is too large to be produced for Al$_2$O$_3$ or Fe particles of any radius. 

Figure~\ref{fig:lx-mie} is computed for spherical particles of a single size, and a distribution of particle sizes must be convolved with the $\omega$ curves for estimates of a non-uniform particle size distribution. The Mie calculations for $\omega$ are also upper limits on $\omega$, since the true $\omega$ of a, e.g., hydrogen-dominated atmosphere containing these species will be somewhat smaller due to gas absorption. Thus, Mie theory can exclude some species and weakly constrain the particle size.

\begin{figure*}
    \centering
    \includegraphics[width=\textwidth]{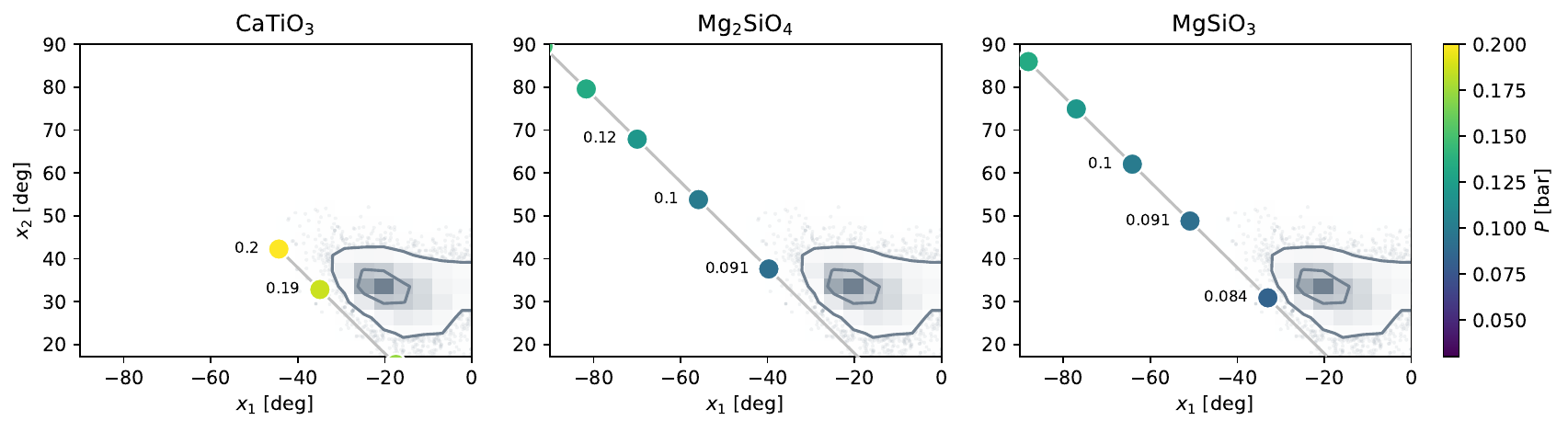}
    \caption{Joint posterior distributions for the planetary longitudes bounding the less reflective region ($x_1,~x_2$) for Kepler-41 b in grayscale, in comparison with the theoretical predictions for several refractory species in each subplot, in color. The gray curves indicate the $1$ and $2\sigma$ contours for $(x_1,~x_2)$. Each species' condensation temperature is converted to a coordinate in $(x_1,~x_2)$, which varies as a function of the cloud pressure, producing the track of colored points which are labeled with the corresponding cloud pressure in units of bar. For example, the observed longitudes spanning the less reflective region of Kepler-41 b are $2\sigma$ consistent with CaTiO$_3$ clouds located near 0.19 bar. Plausible cloud pressures for all species range from 0.08 to 0.2 bar. }
    \label{fig:condensation-longitudes}
\end{figure*}

Figure~\ref{fig:condensation-longitudes} shows another technique for identifying which condensates best match the inhomogeneous phase curves. The reflected light model has two free longitudinal parameters $x_1$ and $x_2$, which indicate the longitudes that bound the low-albedo region of the planet. At longitudes $x_1 < x < x_2$ in the hot substellar region of the dayside, the atmosphere is too hot for condensates to form. These longitudes corresponding to the cloudless-to-cloudy transition are therefore also the longitudes where the local temperatures $T(x_1)$ and $T(x_2)$ drop below the stability curve of the cloud species. The thermal emission phase curve model,  constructed with the $\hml$ basis functions, describes the temperature of the planet at each latitude and longitude, so we may calculate the local temperatures from the best-fit thermal emission map $T(x_1)$ and $T(x_2)$, evaluated at the longitudes of the transition from cloudless-to-cloudy regions $x_1$ and $x_2$ from the reflected light distribution. We can then compare the inferred transition longitudes and their corresponding temperatures to the stability curves of likely condensate species. If the inferred $(x_1, x_2)$ coordinates from the reflected light model are consistent with a species' predicted starting and ending longitudes, we have identified a candidate condensate for that atmosphere.

There is at least one more under-constrained parameter to consider in the calculation outlined above. Temperatures within planetary atmospheres vary with latitude and longitude, as we have encoded with the $\hml$ basis, but also with altitude or pressure, which we have neglected up to this point. Just as the condensation curves of various species are crossed as one moves farther from the sub-stellar point, temperature decreases with increasing altitude and therefore condensation can only occur above some altitudes. So theoretical predictions for the start/stop longitudes of the dark region are themselves functions of pressure, $(x_1(p),~x_2(p))$. For each species, starting and ending longitudes of the dark region are thus compatible with some range of pressures where condensation may occur. The unknown cloud-deck pressure is not a fitting parameter in our models, and there are few theoretical expectations for a reasonable range, so we evaluate a range of pressures for each species.

The three panels of Figure~\ref{fig:condensation-longitudes} show the joint posterior distribution for the starting and ending longitudes of the dark region, $x_1$ and $x_2$ in grayscale. The track of colored points joined by a silver line show the $(x_1(p),~x_2(p))$ coordinates predicted from combining the stability curve of each species with the temperature map for Kepler-41 b. The posterior distribution for $(x_1,x_2)$ is $2\sigma$-consistent with the theoretical predictions for perovskite, forsterite, and enstatite clouds placed at pressures ranging from 0.08 to 0.2 bar. 

In summary, theoretical considerations from Mie theory and condensation maps yield the following key inferences: (1) two of the five most refractory species can be ruled out by their scattering properties; (2) the plausible particle sizes for these condensates is $10^{-2} - 10^{-1}~\mu$m; (3) the three plausible remaining species could condense in the correct locations in the atmosphere to reproduce the inhomogeneity observed in the phase curves; (4) the pressure at the cloud layer may be in the range 0.08 to 0.2 bar.

\subsection{Observational biases}

\subsubsection{Single scattering albedo}

The first, trivial bias implied by parameterizing $\omega$ arises from the intrinsic brightness of the planet. A more reflective planet with larger $\omega$ will produce a larger signal in reflected light and can therefore be detected more easily. 

\begin{figure}
    \centering
    \includegraphics[width=\columnwidth]{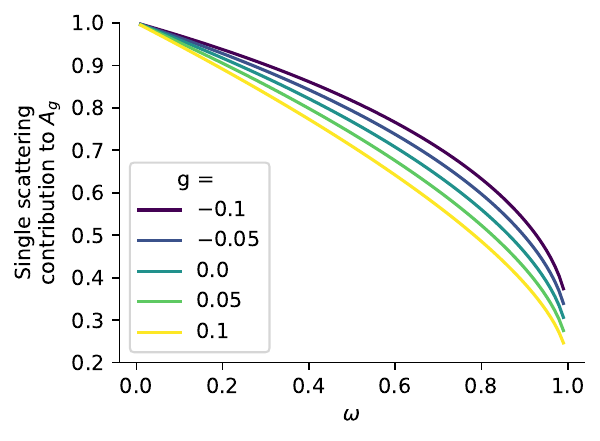}
    \caption{Relative contributions to the geometric albedo from Rayleigh single scattering events and isotropic multiple scattering \citep{Heng2021}.}
    \label{fig:bias_omega}
\end{figure}

A more subtle observational bias arises from the relative contributions of single and multiple scattering for different single scattering albedos. In Figure~\ref{fig:bias_omega}, the curves represent the fraction of the geometric albedo contributed by Cornette-Shanks single scattering, which can be computed trivially with Equations 22 and 24 of \citet[][see also their Figure~2a]{Heng2021}. Single scattering accounts for more than $80\%$ of the geometric albedo of the planet when $\omega < 0.4$ for $g \approx 0$, and the contribution from multiple scattering becomes non-negligible for $\omega > 0.4$.

This suggests a slight limitation of the technique presented here and in \citet{Heng2021}, since the framework can apply any single-scattering phase function to the reflected light phase curve, but the multiple scattering is assumed to be isotropic. Figure~\ref{fig:bias_omega} shows that at large single scattering albedos, a significant fraction of the reflected light phase curve is contributed by isotropic multiple scattering. As a result, we should expect the shape of the phase curve to constrain the single scattering phase function for $\omega < 0.4$; otherwise the assumption of isotropic multiple scattering begins to affect the shape of the reflected light phase curve. This partly explains why we found in Section~\ref{sec:theory_isotropic} that the isotropic multiple scattering assumption has only a very small effect on the reflected light phase curve shape at $\omega = 0.5$.

It is possible that for small enough $\omega$, the single scattering will swamp the multiple scattering signal, but the first-mentioned observational bias against detecting faint objects will make it more difficult to measure the shapes of these phase curves. Therefore, we conclude that the isotropic multiple scattering assumption in the \citet{Heng2021} reflected light model likely has a small effect on this analysis.  

These results are in good agreement with the Monte Carlo radiative transfer simulations of HD 189733 b by \citet{Lee2017}, who found that Rayleigh-like scattering phase functions are necessary where single scattering dominates, and when multiple scattering dominates, isotropic phase functions are a good approximation. 

\subsubsection{Scattering asymmetry parameter}

Another bias affects phase curve scattering parameter inference through the asymmetry parameter $g$. A value of $g < 0$ implies a net back-scattering atmosphere while $g > 0$ indicates predominantly forward scattering. A back-scattering atmosphere will reflect more light near secondary eclipse than an atmosphere which is forward scattering, and therefore the reflected light phase curve amplitude will be larger for planets with back-scattering atmospheres. Forward scattering may be detectable by highly precise observations near transit \citep{GarciaMunoz2018}. 

To demonstrate the effect of this bias on retrieving phase curve scattering parameters, we can inject many reflected light phase curves into random normal noise with standard deviations of 50 ppm, with no thermal emission component from the planet. We solve for the best $\omega$ and $g$ which fit the light curve with the L-BFGS-B minimizer, and plot the true and best-fit values for each parameter in Figure~\ref{fig:bias_g}. The biases presented in this section are each reflected in the variance of the recovered parameters as a function of $\omega$ and $g$. As outlined in the previous subsection, the variance in $\omega$ is largest for small $\omega_\mathrm{true}$, since the phase curve amplitude scales with $\omega_\mathrm{true}$. As $g$ becomes more negative, strong back-scattering amplifies the peak of the phase curve near secondary eclipse and the best-fit $g$ becomes more precise. For $g > 0$, the forward-scattering pushes more flux into the planet's atmosphere, reflecting less to the observer near secondary eclipse, reducing the precision on $g$. The precision on $g$ ranges from $<1\%$ for $g < -0.1$, about 3\% at $g=0$, and greater than $10\%$ for $g > 0.5$.

The observational bias against confident detections of $g > 0$ affects our ability to detect reflected light from large particles. As shown in Figure~\ref{fig:lx-mie}, small particles ($r < 0.01~\mu$m) tend to reflect with $g \rightarrow 0$, regardless of species. The strong forward-scattering nature of larger particles, which greatly reduces the observed flux in the phase curve, implies that we are biased towards detecting reflected light phase curves from planets with clouds made of smaller particles. An interesting exception to this trend is iron, which slightly back-scatters light when the particle radius is near 0.1 $\mu$m.

\section{Discussion} \label{sec:discussion}

\subsection{Correcting the single scattering albedo for Kepler-7 b}

In comparing our results for Kepler-7 b with \citet{Heng2021}, we found that we derive similar values for each fitting parameter except for the single scattering albedo of Kepler-7 b. We note here that we have discovered a typo in the code that produced the corner plot in Figure~4b and the table in Extended Data Figure~6 of \citet{Heng2021}, which incorrectly reported the single scattering albedo for the more reflective region, $\omega^\prime$, and its sum with the albedo less reflective region $\omega \equiv \omega_0 + \omega^\prime$. \citet{Heng2021} reported $\omega = 0.12 \pm 0.05$, whereas the corrected calculation is in fact much closer to unity, $\omega = 0.99 \pm 0.01$. The corrected values are in much better agreement with the analysis by \citet{Hu2015} for example, whose Table~1 shows an analogous quantity to $\omega_0$ and $\omega^\prime$ written as $r_0$ and $r_1$. We list all literature values and our revisions for the albedo of Kepler-7 b in Table~\ref{tab:Kepler7-lit}.

\subsection{Phase curve photometry as a cloud detector}

We have presented evidence for clear and cloudy skies in exoplanet atmospheres based on time-series photometry alone in Section~\ref{sec:results}. Traditionally this inference has been done via spectroscopy, either in emission or transmission. For targets where both observations are available, a combination of photometry and spectroscopy should be used to infer cloudiness, but by precisely testing the shape of a single-band, optical phase curve we can infer cloudiness.

The inhomogeneity inferred from a phase curve is a stronger test of the partial cloudiness of an atmosphere when compared with the effect of the scattering phase function on the shape of the phase curve, as outlined in Section~\ref{sec:scattering-phase-functions}, because we currently have no way of producing the inhomogeneity from a cloud-free atmosphere. Inhomogeneity produces asymmetries in the phase curve that are readily detected with high S/N, \kepler-like observations. Identifying the Rayleigh scattering expected for a clear H$_2$/He atmosphere from the phase curve shape alone is quite challenging since the effect of different scattering phase functions on the phase curve amplitude is small.

\subsection{No cloudy-to-clear transition temperature}

It is tempting to imagine that there is a transition in equilibrium temperatures above which atmospheres are too hot to form condensates and are always clear or cloud-free. In practice, there is likely a transition regime where condensates may form but they do or do not based on more complicated physics than is considered in this work. As more planets are discovered and their phase curve photometry fills in Figure~\ref{fig:homogeneity}, we may find cloud-free, homogeneous atmospheres throughout this temperature range.

\subsection{Condensate species and their sizes}

In Section~\ref{sec:scatterers} we propose a particle size range consistent with the optical properties of the exoplanet atmospheres in reflected light. \citet{LecavelierDesEtangs2008} find a similar particle radius range for enstatite grains in the atmosphere of HD~189733~b, and a similar size range was later invoked by \citet{Pont2013} for the same planet. Enstatite grains have also been invoked to fit interferometric spectroscopy of HD~206893~B \citep{Kammerer2021}. \citet{Wakeford2015} considered transmission spectroscopy of cloudy planets with varying compositions and particle sizes, and found that the spectrum of HD~189733~b could be adequately fit by enstatite clouds with particle sizes ranging from $10^{-2}-10^{-1}~\mu$m, consistent with the interpretation of this work.

\citet{Munoz2015} produced synthetic phase curves for Kepler-7 b as a function of $\omega$ and $g$, assuming a double Henyey-Greenstein scattering phase function, and performed a $\chi^2$ minimization to select the most-likely properties of the scattering condensates. The likely species and particle sizes enumerated in \citet{Munoz2015} are in good agreement with our results from the case study on Kepler-41 b.

\citet{Webber2015} studied reflected light phase curves with inhomogeneous cloud cover applied to Kepler-7~b. The authors find that enstatite or forsterite condensation provides the best match to the observed reflected light asymmetry and albedo for Kepler-7~b. Similarly, \citet{Oreshenko2016} suggest that enstatite or forsterite (plus iron, corundum, or titanium oxide) provide reasonable matches to the phase offset observed for Kepler-7~b. The analysis presented in Section~\ref{sec:scatterers} can rule out some of the species considered in these previous works based on the scattering properties $(\omega, g)$ of the condensates.

\citet{Parmentier2016} used GCMs, condensation temperatures, and scattering properties of candidate condensates to constrain ensemble cloud properties of hot Jupiters. In the equilibrium temperature range of planets discussed in this work, the authors highlight silicate and perovskite clouds for cooler atmospheres (Kepler-7 b and Kepler-41 b, respectively) and corundum clouds for hotter atmospheres (HAT-P-7 b) as potential scatterers consistent with phase curve observations. Their simulated planets with the highest albedos and largest reflected light asymmetries are produced for particles with sizes near $10^{-1}~\mu$m, which is consistent with the size upper limits implied by the small scattering asymmetries in this work.

Our results are also consistent with \citet{Lee2016, Lee2017}, who presented 3D radiative-hydrodynamic simulations with a kinetic, microphysical, non-equilibrium mineral cloud model for HD 189733 b. Their condensate particle sizes are typically sub-micron, with cloud formation at the western limb and at mid-latitudes dominated by enstatite and forsterite.

Cloud microphysics models by \citet{Powell2019} suggest that the west limbs of hot Jupiters with $T_\mathrm{eq}=2000$ K may have, for example, forsterite condensation with number densities which peak near $10^{-2}$ and 1 $\mu$m (see their Figure~5). These particle sizes are consistent with the observations presented here, in the interpretation in Section~\ref{sec:scatterers}. More generally, \citet{Gao2020} provide observational and theoretical evidence that silicates are the dominant condensates in hot atmospheres with $T_\mathrm{eq} \gtrsim 1600$ K, which includes all planets in this sample.

\subsection{Variability in exoplanet atmospheres}

Uncertainty in time-series detrending should inform posterior inferences. The impact of the detrending techniques on the inferred phase curve parameters is a key challenge in interpreting reflected light phase curves in the literature. Often, conservative detrending approaches are applied independently from the posterior inference, producing, for example, phase curves that appear highly variable with time \citep{Armstrong2016}; while more liberal detrending risks removing signals of interest. Theoretically, there are few or no predictions for large variations in the thermal emission components of the phase curve on year-long timescales. \citet{Komacek2020} used general circulation models to show that the variations in eclipse depth, phase curve amplitude, and hotspot offset should all change by a few percent or less. High resolution pseudospectral simulations by \citet{Y-K.Cho2021} are also in agreement with this amplitude of variations in the thermal emission. Larger variations might be expected in reflected light since, for example, there may be significant albedo variations driven by partial cloud cover, as we see on planets in the Solar System. The observations presented in this work do not support larger than a few percent in the geometric albedo of hot Jupiters on year-long timescales.

Claims of variations of exoplanet atmospheric properties should be approached with caution. Many observations support the theoretical expectations against significant thermal variability: \citet{Agol2010, Kilpatrick2020} found that the thermal flux of HD 189733 b and HD 209458 b change by less than a few percent between eclipses. \citet{Jones2022} found a similar upper limit on the optical flux variations of the ultra-hot Jupiter KELT-9 b. Recent work by \citet{Lally2022} showed that apparent variations like those invoked for the \kepler phase curve of HAT-P-7 b may be explained by uncorrected stellar or instrumental artifacts.

\section{Conclusion} \label{sec:conclusion}

\begin{itemize}
    \item We have introduced a Bayesian inference framework for inferring optical properties of exoplanet atmospheres from phase curves, which includes flux contributions from reflected light from a potentially inhomogeneous atmosphere, thermal emission, ellipsoidal variations, Doppler beaming, and stellar rotation (Section~\ref{sec:methods}).
    \item We applied this inference framework to the \kepler phase curves of five hot Jupiters to measure atmospheric homogeneity and time-variability, as well as an investigation into the scattering properties which constrain the likely condensates in inhomogeneous atmospheres. We use the Leave-one-out cross-validation technique with Bayesian stacking for model selection (Section~\ref{sec:loo}-\ref{sec:modelselection}).
    \item The observations suggest an inhomogeneous albedo distribution for three of the five planets (Figure~\ref{fig:homogeneity}), which we interpret as asymmetric cloud distributions.
    \item We enumerate the phase curve parameters of each planet (Table~\ref{tab:params}), and show the single scattering albedos for the reflective region of each planet (Figure~\ref{fig:omegas})
    \item None of the planets exhibit significant geometric albedo variations in time (Figure~\ref{fig:yearly-A_g}).
    \item For Kepler-41 b, we have identified perovskite, forsterite, and enstatite as possible scattering species consistent with the reflected light phase curves using condensate stability curves (Figure~\ref{fig:hat7-tp}) and Mie theory (Figure~\ref{fig:lx-mie}). The condensate particle radii may be in the range $10^{-2}-10^{-1}~\mu$m, and the pressure at the clouds may be in the range $0.08-0.2$ bar (Figure~\ref{fig:condensation-longitudes}).
    \item We demonstrated that analytic phase curves with isotropic multiple scattering are in excellent agreement with full Rayleigh multiple scattering calculations (Figure~\ref{fig:phase-curve-comparison}).
\end{itemize}

\begin{acknowledgements}
    We are grateful to our referee, Nicolas Cowan. We gratefully acknowledge the open source software which made this work possible: \texttt{astropy} \citep{astropy:2013, astropy:2018, astropy:2022}, \texttt{ipython} \citep{ipython}, \texttt{numpy} \citep{numpy}, \texttt{scipy} \citep{scipy}, \texttt{matplotlib} \citep{matplotlib}, \texttt{JAX} \citep{jax2018github}, \texttt{arviz} \citep{arviz_2019}, \texttt{numpyro} \citep{numpyro}, \textsc{FastChem} \citep{Stock2018MNRAS.479..865S, Stock2022, Kitzmann2023arXiv230902337K}, \texttt{LX-MIE} \citep{Kitzmann2018a}, \texttt{celerite2} \citep{celerite1, celerite2} \texttt{exoplanet} \citep{exoplanet:joss}, \texttt{lightkurve} \citep{LightkurveCollaboration2018}, \texttt{corner} \citep{Foreman-Mackey2016}, \texttt{kelp} \citep{Morris2021_hml}. This research has made use of the SVO Filter Profile Service (\url{http://svo2.cab.inta-csic.es/theory/fps/}) supported from the Spanish MINECO through grant AYA2017-84089. This research has made use of the NASA Exoplanet Archive, which is operated by the California Institute of Technology, under contract with the National Aeronautics and Space Administration under the Exoplanet Exploration Program.
\end{acknowledgements}

\bibliographystyle{aa} 
\bibliography{bibliography}

\appendix

\section{Index of symbols}

Table~\ref{tab:symbols} describes the symbols used in this work and their definitions.

\begin{table*}
\centering
\setstretch{1.3}
\begin{tabular}{llp{10cm}}
    {\bf Symbol} & {\bf Name} & {\bf Description} \\ \hline
     $a/R_p$ & Semimajor axis & Orbital semimajor axis normalized by the planetary radius \\
     $A_g$ & Geometric albedo & \\
     $\alpha$       & Dimensionless fluid number & From \citet{Morris2021_hml}\\
     $\alpha_\mathrm{ellip}, \alpha_\mathrm{Doppler}$ & Amplitude coefficients & Factors of order unity which parameterize the amplitude of ellipsoidal variations and Doppler beaming, respectively\\
     $C_{11}$ & Spherical harmonic power & Power in the lowest-order spherical harmonic coefficient, which describes the day-night temperature contrast in the \hml basis \\
     $\Delta\phi$   & Hotspot offset & This parameter approaches the longitudinal offset of the hotspot from the substellar point in the limit that $\omega_\mathrm{drag} \gg 1$, where in this work usually $\omega_\mathrm{drag} = 4.5$\\
     $\delta_F$ & Dilution & Flux dilution factor due to neighboring stars within aperture\\
     $g$ & Scattering asymmetry parameter & Varies on $-1 < g < 1$ from pure back scattering to pure forward scattering\\
     $f$ & Greenhouse parameter & From \citet{Morris2021_hml}\\
     $I(\theta, \phi)$ & Stellar/planetary intensity & Evaluated for a given surface location from \citet{Morris2021_hml}\\
     $P$ & Orbital period & Planetary orbital period\\
     $P_\mathrm{rot}$ & Rotation period & Stellar orbital period\\
     $R_p/R_\star$  & Radius ratio & Ratio of planet to stellar radius \\
     $\omega$       & Single-scattering albedo & Fraction of light scattered in single scattering event \citep{Heng2021}\\
     $\omega_0$       & Minimum single-scattering albedo & Fraction of light scattered in single scattering event from the less-reflective surface of the inhomogeneous atmosphere model \citep{Heng2021}\\
     $\omega_\mathrm{drag}$ & Dimensionless drag parameter & From \citet{Morris2021_hml}\\
     $\theta$       & Planetary colatitude & Planetary colatitude from ($0, \pi$) \\
     $\phi$         & Planetary longitude & Planetary latitude from ($0, 2\pi$) \\
     $\phi_\mathrm{orb}$ & Orbital phase & Normalized from ($0, 1$) where 0.5 is secondary eclipse\\ 
     $\xi$          & Orbital phase & Normalized from ($-\pi, \pi$) where zero is secondary eclipse\\
     $\mathcal{F}_\lambda$ & Response function & Detector/filter response as a function of wavelength\\
     $\mathcal{B}_\lambda$ & Planck function & \citet{Planck1901}\\
     $\lambda^e$ & Occultation model & \citet{exoplanet:agol20} star/planet occultation model\\
     $F_p/F_\star$ & Observed flux ratio & Ratio of planetary to stellar flux\\
     $\Psi$ & Integral phase function & From \citet{Heng2021} \\
     $q$   & Phase integral & From \citet{Heng2021} \\
     $x_1$ & Start longitude & Start longitude of the less-reflective region from \citet{Heng2021}\\
     $x_2$ & Stop longitude & Stop longitude of the less-reflective region from \citet{Heng2021}\\
     $x_1^\prime$ & Primed start longitude & Reparameterization of the start longitude: $\sin(x_1)$\\
     $x_2^\prime$ & Primed stop longitude & Reparameterization of the stop longitude: $\sin(x_2)$\\
     $T(\theta, \phi)$ & Temperature map & Map of the temperature as a function of latitude and longitude\\
     $T_{\rm d, n}$ & Hemispheric temperature & Dayside and nightside integrated hemispheric temperature\\
     $\mathcal{N}(\mu, \sigma)$ & Normal distribution & Gaussian with mean $\mu$ and standard deviation $\sigma$\\
     $\mathcal{U}$ & Uniform distribution & \\
     $\sigma_\mathrm{GP}$ & GP std. dev. & Standard deviation (amplitude) of the Gaussian process\\
     $\rho_\mathrm{GP}$ & GP timescale & Typical timescale of oscillations in the Mat{\'e}rn 3/2 kernel\\
\end{tabular}
\caption{Definitions of symbols used in this work.}
\label{tab:symbols}
\end{table*}

\section{Biases}

\begin{figure*}
    \centering
    \includegraphics[width=\textwidth]{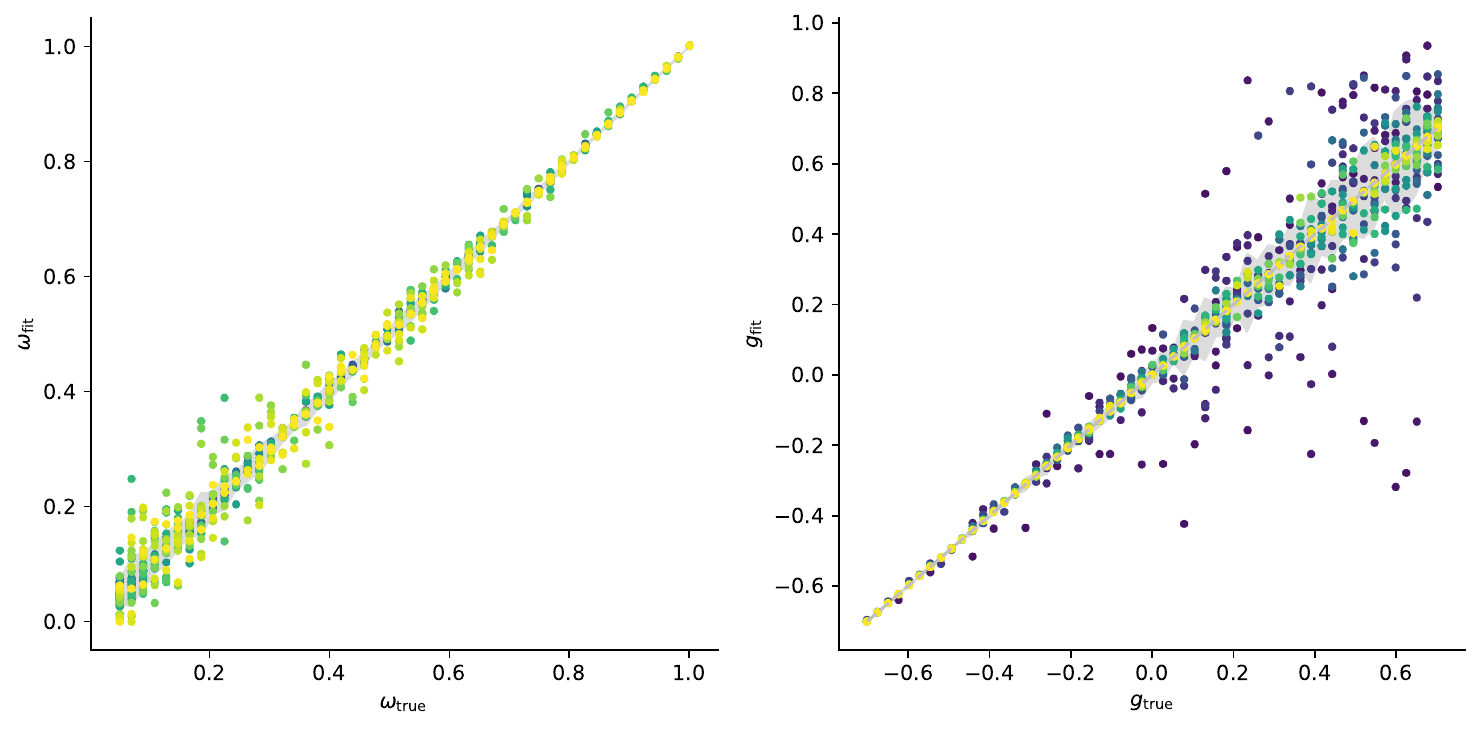}
    \caption{Comparison of the true (injected) single scattering albedo $\omega$ and scattering asymmetry parameter $g$ with their best-fit values in simulated light curves of reflected light phase curves, with no thermal emission and Gaussian noise. As $\omega_\mathrm{true} \rightarrow 0$, the reflected light signal diminishes, reducing the precision in $\omega_\mathrm{fit}$. For $g > 0$, the planet's atmosphere scatters more light into the atmosphere than back at the observer, reducing the reflected light signal, and reducing the precision on $g_\mathrm{fit}$. The color of each point in the $g$ panel denotes the value of $\omega$ for that point, and vice versa. Note that $g$ is least precise for the smallest $\omega$ values (purple).}
    \label{fig:bias_g}
\end{figure*}

An illustrative battery of phase curve injection/recovery tests are shown in Figure~\ref{fig:bias_g}. These exercises demonstrate the biases which affect retrieval of the single scattering albedo $\omega$ and scattering asymmetry parameter $g$. 

\section{Results for Kepler-7 b}

There are many results for the geometric albedo in the literature, which we collate in Table~\ref{tab:Kepler7-lit}.

\begin{table*}
    \centering
    \setstretch{1.5}
    \begin{tabular}{r|ccc}
        & $A_g$ & $\omega_0$ & $\omega^\prime$ \\ \hline
    \citet{Demory2011b} & $0.32 \pm 0.03$ & -- & --\\
    \citet{Demory2013} & $0.35 \pm 0.02$ & -- & --\\
    \citet{Esteves2015} & $0.248^{+0.071}_{-0.073}$ & -- & -- \\
    \citet{Hu2015} & $0.28 \pm 0.006$ & $<0.072$ & $0.92$\\
    \citet{Munoz2015} & $0.2-0.3$ & & $\sim 1$\\
    \citet{Heng2021} & $0.25^{+0.01}_{-0.02}$ & $0.0136^{+0.0132}_{-0.0093}$ & {\color{red} $0.115^{+0.044}_{-0.049}$}\\
    This work & ${0.27}_{-0.01}^{+0.01}$ & $0$ (fixed) & ${0.99}_{-0.01}^{+0.00}$ \\
    \end{tabular}
    \caption{Comparison of literature results on the albedo of Kepler-7 b.}
    \label{tab:Kepler7-lit}
\end{table*}

\end{document}